\documentclass[reprint,
	nofootinbib,
	amsmath,
 	amssymb,
	amsfonts,
 	pra,
]{revtex4-1}

\usepackage[utf8]{inputenc} 
\usepackage[T1]{fontenc}    
\usepackage[brazil]{babel}
\usepackage{graphicx}
\usepackage{hyperref}       
\usepackage{url}            
\usepackage{booktabs}       
\usepackage{amsfonts}       
\usepackage{amsmath}
\usepackage{amssymb}
\usepackage{nicefrac}      		
\usepackage{xcolor}
\DeclareMathOperator{\arccosh}{arccosh}

\begin{document}

\title{Uma introdução à evolução do Universo segundo sua geometria e composição \\
\small{-- An introduction to the time evolution of model universes -- }}

\author{Vinicius~S.~Aderaldo}
\email{vinicius.aderaldo@ufpel.edu.br}
\author{Victor~P.~Gonçalves}
\email{barros@ufpel.edu.br}
\affiliation{Instituto de Física e Matemática, Universidade Federal de Pelotas,  Pelotas, RS, Brasil}

\date{\today}

\begin{abstract}
Neste artigo apresentamos uma introdução à evolução do Universo predita pela equação de Friedmann, a qual leva em consideração a composição e geometria do Universo. A dependência da solução da equação de Friedmann na geometria é analisada, assim como a sua solução para diferentes  combinações para os   constituintes básicos que formam o Universo. Os distintos  comportamentos possíveis da evolução temporal do Universo são determinados e o cenário predito pela Cosmologia Padrão é apresentado. 
\\
\textbf{Palavras-Chave:}~Relatividade Geral, Cosmologia, Universo, Expansão.\\

In this paper we describe the evolution of the Universe in terms of the Friedmann equation, which takes into account of the composition and geometry of the Universe. The dependence of the solution on the geometry and composition for different combinations of the basic constituents are discussed. The distinct behaviours for the temporal evolution of the Universe are determined and the scenario predicted by the Standard Cosmology is presented.
\end{abstract}
                              
\maketitle

\section{Introdução}
\label{introducao}
Quando estudamos o Universo em larga escala estamos, de fato, estudando Cosmologia. Etimologicamente, Cosmologia advém do grego $\kappa\textrm{ó}\sigma\mu\textrm{o}\varsigma$, "cosmos", significando~"ordem"~e~$\lambda\textrm{o}\gamma\acute{\iota}\alpha$, "logia", significando~"estudo". Nas palavras de um dos três laureados com o Prêmio Nobel de Física de 2019, P.~J.~E.~Peebles (1935--): "A cosmologia física é a tentativa de dar sentido à natureza de larga escala do mundo material ao nosso redor, através dos métodos das ciências naturais."~\cite[p.~3]{peebles}. Para as grandes distâncias presentes na descrição do Cosmos, {denominadas escalas cosmológicas ($\geq100\textrm{Mpc}$\footnote{$1\textrm{pc}=3,086\times10^{16}\textrm{m}$~\cite{liddle}.}), o Princípio Cosmológico\footnote{Ver Seção~\ref{principio_cosmologico}.} é válido e }a interação dominante é a interação gravitacional, uma vez que em escalas suficientemente grandes o Universo é eletricamente neutro, o campo magnético negligenciável e as forças nucleares fraca e forte totalmente desprezíveis~\cite{barbara}. {Sendo assim, o Universo em  escalas cosmológicas} é descrito pela teoria da Gravitação.

{No início da modernidade, a teoria da gravitação de Isaac Newton (1642--1727)  apresentada em 1687  em sua obra \textit{Philosophiae Naturalis Principia Mathematica} (Princípios Matemáticos da Filosofia Natural)  tornou-se a mais bem-sucedida teoria da Gravitação}. A teoria da gravitação de Newton explica muito bem, e.g., o movimento da Lua em torno da Terra e as leis de Kepler~\cite{acioli,roos}. No entanto, tal teoria é limitada, falhando em alguns aspectos. Por exemplo, em se tratando de uma força que se manifesta perante uma interação entre corpos dotados de massa, uma questão nada trivial diz respeito à natureza dessa força.  Um outro problema reside no fato de que, sendo uma força com dependência no inverso do quadrado da distância e com o produto das massas dos corpos envolvidos na interação, não temos dependência temporal alguma. Assim, essa força se manifesta de forma instantânea, violando o {o que ficaria conhecido como o} segundo postulado da Relatividade Restrita~\cite{cheng2}. Tais aspectos motivaram Albert Einstein (1879--1955) a desenvolver a teoria da Relatividade Geral~\cite{einstein1915}, a qual é considerada {uma} teoria clássica da Gravitação. Portanto, as equações básicas que regem o Universo em larga escala e, consequentemente, a Cosmologia, são as equações de campo propostas por Einstein para a teoria da Relatividade Geral, que vinculam geometria e constituição do Universo (Seção~\ref{RG_EF}).

{No início dos anos de 1920, Alexander Friedmann (1888--1925)~\cite{friedmann1922,friedmann1924} e Georges Lemaître (1894--1966)~\cite{lemaitre2,lemaitre} obtiveram, independentemente, soluções não estáticas para as equações de campo propostas por Einstein~\footnote{A contribuição de Friedmann para a Cosmologia é discutida em detalhes na referência.~\cite{ioav2}.}. Entretanto, foi no } ano de 1929,  que ocorreu a descoberta que mudaria o paradigma  de um Universo estático vigente até então~\footnote{Uma discussão detalhada acerca do modelo estático de Einstein bem como de sua instabilidade pode ser obtida na referência~\cite{soares}.}. Tal feito é devido ao astrônomo Edwin P.~Hubble~(1889--1953) que descobriu que o Universo está se expandindo~{\cite{hubble,hubble2}}. Para que isso fosse possível, Hubble fez uso de descobertas anteriores: a da astrônoma Henrietta S.~Leavitt (1868--1921) no que diz respeito às velas padrão\footnote{{Objeto cuja luminosidade é bem conhecida, servindo dessa maneira como "régua cósmica".}}~\cite{henrietta} e a do astrônomo Vesto M.~Slipher (1875--1969) devido aos seus estudos acerca dos desvios espectrais~\cite{slipher,slipher2}\footnote{Para uma discussão mais detalhada acerca da descoberta de Hubble, veja a referência~\cite{ioav}.}. Posteriormente, em 1998, os grupos de pesquisadores associados ao \textit{The Supernova Cosmology Project}~\cite{perlmutter,perlmutter2} e \textit{High-z Supernova Search Team}~\cite{riess} verificaram de forma independente que, na verdade, a expansão do Universo verificada por Hubble é acelerada. A descrição e compreensão deste resultado é um dos grandes desafios para a Cosmologia. 


A formulação de uma teoria para a Cosmologia ainda é tema de intenso debate. Neste trabalho iremos focar em apresentar os fundamentos e  implicações do  {{modelo $\mathbf{\Lambda}$CDM}~\footnote{{Proveniente de: $\Lambda$ {\it{Cold Dark Matter}}, onde $\Lambda$ é a Constante Cosmológica}. Ver Seção~\ref{cosmologia_padrao}.}, o qual é atualmente considerado o modelo padrão da Cosmologia, pois fornece o melhor ajuste aos dados experimentais existentes~\cite{pdg2020}}. Tal teoria fundamenta-se fortemente no Princípio Cosmológico, o qual nos diz que, em escalas cosmológicas, podemos considerar o Universo como sendo homogêneo e isotrópico, isto é, que o Universo possui seu conteúdo material distribuído igualmente por toda a sua extensão, além de não existir uma direção preferencial para descrevermos os eventos físicos presentes nele. {O Princípio Cosmológico implica que todas as quantidades observáveis serão invariantes por translação (homogeneidade) e rotação (isotropia). Além disso, quando} assumimos o Princípio Cosmológico, as equações propostas por Einstein para a Relatividade Geral recaem, naturalmente, na chamada equação de Friedmann~\cite{friedmann1922,friedmann1924}. Tal equação descreve a evolução temporal do Universo e, portanto, é uma das equações mais importantes da Cosmologia. Neste artigo iremos apresentar a sua derivação e descrever suas soluções para distintas geometrias e diferentes combinações de constituintes básicos. Por fim, trataremos da solução para um Universo plano composto por três constituintes básicos, a qual representa a visão aceita atualmente pela Cosmologia Padrão.

Este artigo tem como foco os alunos de graduação, visando situá-los no cenário atual da Cosmologia bem como os apresentar às equações que formam a sua estrutura. {Seguiremos uma abordagem semelhante àquela presente nas Refs. \cite{barbara,bari}, buscando apresentar em maior detalhe as derivações das equações e discussão dos conceitos, assim como iremos utilizar em nossa analise os valores mais recentes dos parâmetros cosmológicos apresentados na última edição do {\it Review of Particle Physics}~\cite{pdg2020}}. Dessa forma, iremos partir das equações de campo propostas por Einstein como sendo a base de toda a descrição do Universo e, através de considerações cabíveis, a saber, o Princípio Cosmológico, obter a equação primordial para nossos estudos, a equação de Friedmann. Feito isso, utilizando a Cosmologia Padrão, iremos explorar as implicações de diferentes considerações no que diz respeito à composição do Universo e à geometria do espaço, verificando o comportamento da evolução do Universo através das respectivas soluções da equação de Friedmann. Através dessas considerações, iremos determinar os intervalos temporais nos quais cada uma das componentes se mostra dominante sobre as demais. Tal descrição nos possibilita não somente compreender o comportamento do Universo ao longo do tempo, a fim de predizer possíveis destinos, mas também calcular quantidades de interesse, como, por exemplo, a idade do Universo. 

Este trabalho está organizado da seguinte forma. Na próxima Seção apresentaremos os fundamentos da Cosmologia Padrão, com ênfase nas equações de Friedmann. Na Seção \ref{curvatura}, {determinaremos o efeito que as distintas curvaturas possíveis para o Universo acarretam em sua evolução}. Assumindo um Universo plano, nas Seções \ref{single_component}, \ref{duas_componentes} e \ref{completo} apresentaremos, respectivamente, as soluções da equação de Friedmann para um Universo composto por um único constituinte, uma combinação de dois constituintes e, por fim, por três constituintes. {Nossas conclusões} são apresentadas na Seção \ref{conclusao}. {Por fim, no material suplementar apresentaremos  a derivação da Equação de Friedmann (Apêndice A) e da equação de fluido (Apêndice B), assim como iremos apresentar uma estimativa para a idade de um Universo caracterizado por máteria  e  termo de curvatura não nulo (Apêndice C).}
\section{Cosmologia Padrão}
\label{cosmologia_padrao}
Para o desenvolvimento dos nossos estudos iremos  considerar o {\textbf{modelo $\mathbf{\Lambda}$CDM}, o qual melhor descreve os atuais dados observacionais ~\cite{pdg2020} e por isso é usualmente  denominando Modelo Padrão da Cosmologia ou {\textbf{Cosmologia Padrão}~\cite{pordeus,froes}}}. Esse modelo leva em consideração as proposições do \textit{Big Bang}, ou seja, que o Universo teve um início extremamente denso, quente e compacto, seguido de um período inflacionário e uma posterior expansão. Sendo assim, muitos também o chamam como \textit{Hot Big Bang Model}~\cite{barbara,froes}. A Cosmologia Padrão leva em consideração o Princípio Cosmológico e como composição atual os seguintes constituintes: matéria bariônica, matéria escura e energia escura. Como desejamos descrever a evolução do Universo não apenas nos dias atuais, mas também para tempos primordiais, é importante que levemos em consideração a componente referente à radiação. Nesta Seção iremos discutir o Princípio Cosmológico e suas implicações, bem como das componentes citadas anteriormente.
\subsection{O Princípio Cosmológico}
\label{principio_cosmologico}
{O Princípio da Relatividade proposto por Einstein para a Relatividade Restrita~\cite{relatividade_book} nos diz que as leis da Física são independentes do sistema de referência inercial adotado. Sendo assim, tais leis devem ser expressas da mesma forma em todos os referências inercias (Covariância das Leis da Física), sendo as mesmas para todos os observadores. Em extensão a esse postulado, Edward A. Milne (1896--1950)  cunhou, em meados dos anos 1930, o \textbf{Princípio Cosmológico} da forma como perdura até os dias de hoje~\cite{milne1,milne2,ronaldo}. A extensão reside no fato de que, além da forma com que descrevemos as leis da Física, a estrutura do Universo deve ser a mesma para observadores em diferentes referenciais  em relação ao fluido cósmico~\cite{milne1,milne2,ronaldo}. Em suma, este princípio nos diz que em escalas apropriadas o nosso Universo é homogêneo e isotrópico.} 

Se analisarmos regiões do Universo da ordem de $\sim1~\textrm{ano-luz}$ podemos observar aglomerações de estrelas, enquanto que em regiões da ordem de $\sim10^{6}~\textrm{anos-luz}$ observamos aglomerações de galáxias, por sua vez em regiões da ordem de $\sim10^{7}~\textrm{anos-luz}$ vemos aglomerações de aglomerados de galáxias~\cite{kip}. No entanto, quando analisamos regiões do Universo da ordem de escalas cosmológicas, i.e. $\geq100\textrm{Mpc}$ ($1\textrm{pc}=3,261~\textrm{anos-luz}$), podemos observar que o Universo é idêntico em qualquer localidade, a distribuição espacial da composição é a mesma, e dizemos que o Universo é homogêneo~{\cite{barbara,Wu_1999,Scrimgeour_2012}}. Por outro lado, em pequenas escalas é importante que haja inomogeneidade, caso contrário não teríamos a formação de sistemas\footnote{\label{densidadenota}Para se ter uma noção da inomogeneidade em pequenas escalas, se considerarmos uma esfera de $3\textrm{m}$ de diâmetro ao nosso redor, teremos uma densidade média de $\sim100\textrm{kg\,m}^{-3}$, enquanto que a densidade média do Universo é da ordem de $\sim10^{-27}\textrm{kg\,m}^{-3}$~\cite{barbara}.}. Veja, por exemplo, o caso do Sistema Solar. Claramente não temos uma homogeneidade, dado que temos planetas com massas distintas e com distâncias distintas entre si. Além disso, temos que há uma direção preferencial na interação Sol-Planeta. Isso nos leva às definições de isotropia e anisotropia. Como temos uma direção preferencial na interação gravitacional entre o Sol e os planetas no caso do Sistema Solar, dizemos que esse sistema é anisotrópico. {É importante enfatizar que se tivéssemos isotropia, não seria possível manter os planetas em órbita ~\cite{barbara}. No entanto, em escalas cosmológicas, o nosso Universo é isotrópico,  tendo a mesma aparência em todas as direções, não existindo uma direção preferencial para as leis da Física. }

\begin{figure}[t]
\centering
\begin{minipage}{8.5cm}
\centering
\includegraphics[width=8.5cm]{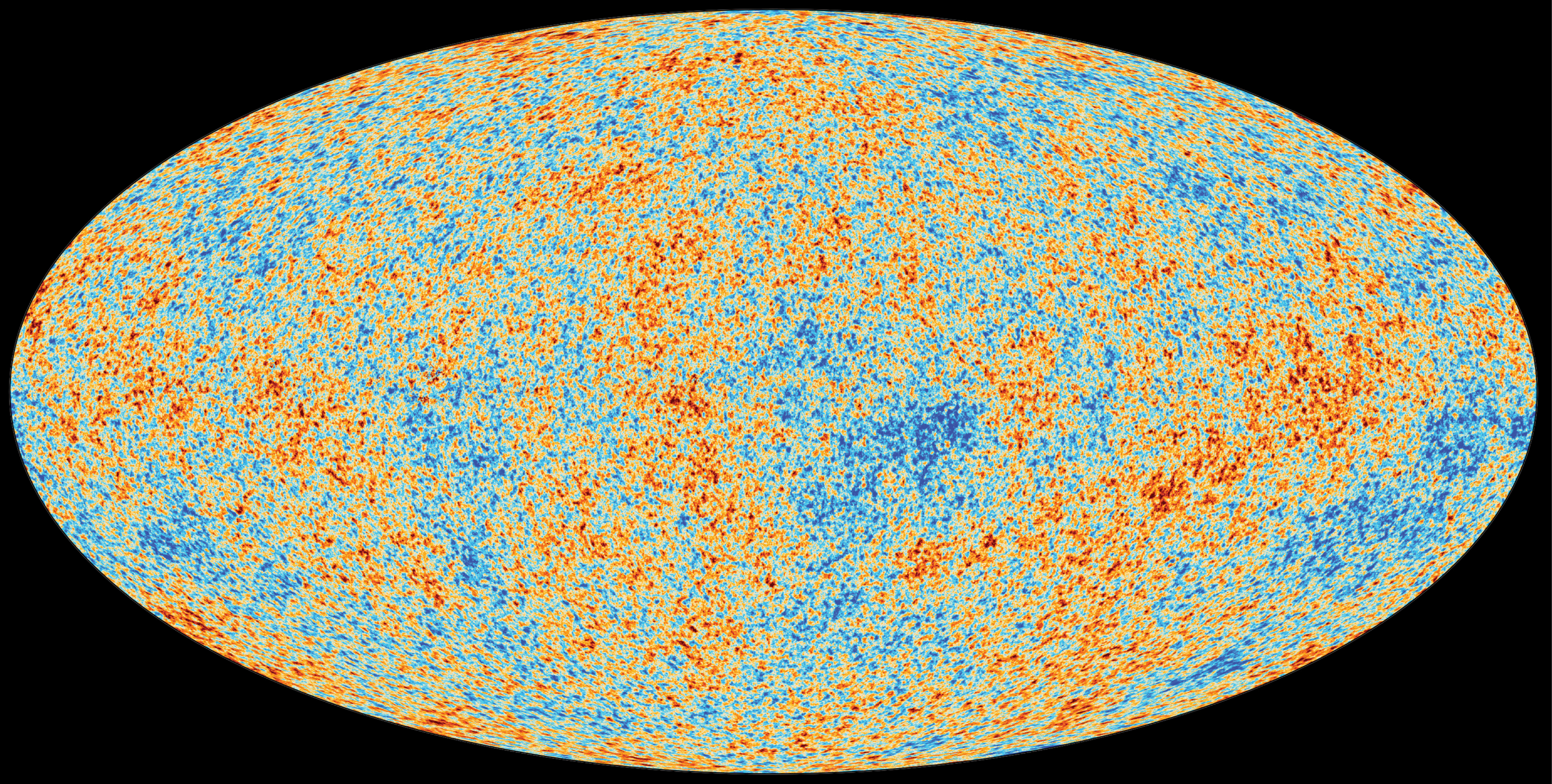}\\
\caption{Radiação Cósmica de Fundo, obtida a partir de dados coletados e distribuídos pelo \textit{Planck Legacy}.}
Créditos:~ESA/Planck Collaboration~\cite{planck}.
\label{CMB}
\end{minipage}
\end{figure}

Observações astronômicas demonstram que de fato, em larga escala, o Universo é homogêneo e isotrópico~{\cite{Wu_1999,Scrimgeour_2012}}. Uma das mais importantes descobertas é concernente à Radiação Cósmica de Fundo\footnote{A Radiação Cósmica de Fundo foi primeiramente detectada pelos astrônomos Arno Penzias (1933--) e Robert Wilson (1936--) ~\cite{penzias}.} que demonstrou possuir uma homogeneidade de uma parte em $10^{5}$ para qualquer direção~\cite{cheng2}. Podemos visualizar a homogeneidade a partir dos dados para a Radiação Cósmica de Fundo coletados e distribuídos pelo satélite \textit{Planck} na \autoref{CMB}, onde vemos as flutuações de temperatura, através das diferentes cores, que representam as "sementes" ~que posteriormente formariam as galáxias e estrelas.

\begin{figure}[t]
\centering
\begin{minipage}{8.5cm}
\centering
\includegraphics[width=8.5cm]{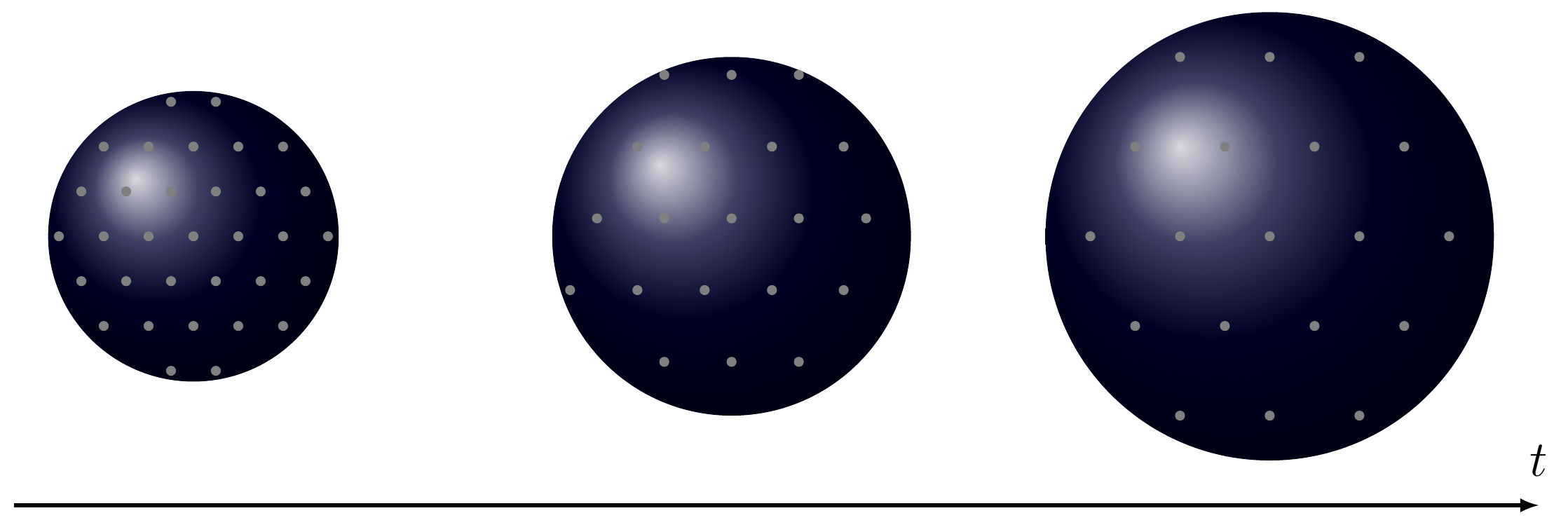}
\caption{Ilustração de uma bola de vinil sendo inflada em função do tempo, onde os pontos cinza representam as moedas-galáxia distribuídas uniformemente sobre a superfície da bola.}
\label{bexiga}
\end{minipage}
\end{figure}

O fato de ser homogêneo e isotrópico implica que não podemos estabelecer um centro, absoluto, para o Universo. Sendo assim, qualquer observador em qualquer ponto do Universo verá a variação das distâncias entre as galáxias da mesma forma~\cite{barbara}. Para compreendermos isso, façamos o seguinte experimento mental. Imagine uma bola de vinil, com algumas moedas (representando as galáxias) sobre a sua superfície e distribuídas de forma uniforme (representando a homogeneidade do Universo). Se começarmos a encher a bola com ar, as moedas se afastarão umas das outras (veja a \autoref{bexiga}). Agora imagine que possamos habitar alguma dessas moedas-galáxia. Com isso, podemos olhar para qualquer direção da bola e constatar o mesmo comportamento das moedas-galáxia vizinhas (representando a isotropia). Note que não podemos estabelecer um centro absoluto, visto que a distribuição de moedas em torno da moeda habitada é esfericamente simétrica. Dessa forma, como não existe um centro absoluto, mas uma variação homogênea e isotrópica das distâncias, podemos considerar um sistema de coordenadas onde a composição do Universo permanece sempre em repouso, chamado \textit{comoving coordinates}~\cite{fayyazuddin}, ou seja, nesse sistema as coordenadas permanecem constantes durante o transcorrer do tempo. Por fim, outra implicação do Princípio Cosmológico diz respeito ao fato de que podemos estabelecer um tempo universal, isto é, se considerarmos diversos observadores no sistema \textit{comoving coordinates} (os chamados \textit{comoving observers}), estes poderão sincronizar os seus respectivos relógios\footnote{Por razões lógicas, consideramos o "tempo zero"~no \textit{Big Bang}.} visto que observarão as distâncias variando da mesma forma e, consequentemente, os eventos ocorrendo da mesma forma~\cite{fayyazuddin,landau}.

Quando assumimos o Princípio Cosmológico, as equações propostas por Einstein para a Relatividade Geral recaem, naturalmente, na chamada equação de Friedmann~\cite{friedmann1922,friedmann1924}, como explicitamente demonstrado no Apêndice A do material suplementar. Tal equação descreve a evolução temporal do Universo e, portanto, é uma das equações mais importantes da Cosmologia. Logo, é interessante fazermos uma breve discussão a respeito da implicação do Princípio Cosmológico no que diz respeito à Relatividade Geral.
\subsection{As Equações de Campo e a Equação de Friedmann}
\label{RG_EF}
As equações da Relatividade Geral configuram o conjunto de equações principais para descrevermos o Universo em larga escala. Tendo em vista a descoberta de uma expansão acelerada~\cite{perlmutter,perlmutter2,riess} e que, como veremos adiante, um Universo contendo apenas matéria e radiação não descreve tal aceleração, devemos levar em consideração um termo cosmológico que proporcione o observado. As equações de campo, já com a Constante Cosmológica $\Lambda$, são escritas como \cite{peebles,barbara}
\begin{equation}
\label{eqdecampo}
R_{\mu\nu}=\frac{8\pi G}{c^{4}}\left(T_{\mu\nu}-\frac{1}{2}g_{\mu\nu}T\right)-\frac{\Lambda}{c^{2}}g_{\mu\nu}\;,
\end{equation}
onde $G$ é a constante gravitacional\footnote{$G=6,673\times10^{-11}\textrm{m}^{3}\textrm{kg}^{-1}\textrm{s}^{-2}$.}. Na Eq.~(\ref{eqdecampo}), temos que o lado esquerdo, composto pelo chamado tensor de Ricci, $R_{\mu\nu}$, descreve a geometria do espaço-tempo. Já no lado direito, temos os termos referentes à composição do Universo, representados pelo tensor energia-momento, $T_{\mu\nu}$. Esse tensor é responsável por descrever a distribuição de matéria e energia tal que, para um Universo homogêneo e isotrópico, podemos considerar como sendo a de um fluido perfeito. Além disso, $g_{\mu\nu}$ é o tensor métrico, que nos diz qual é a relação entre as distâncias espaço-temporais e os intervalos das coordenadas. Por fim, $T$ é o escalar energia-momento, que é obtido através da contração $T=T_{\mu\nu}g^{\mu\nu}$\footnote{Perceba que estamos usando a convenção de Einstein para somatórios, onde fica implícita a soma sobre os índices repetidos.}, e $\Lambda$ é a já citada Constante Cosmológica. 

\begin{figure}[t]
\centering
\begin{minipage}{8.5cm}
\centering
\includegraphics[width=8.5cm]{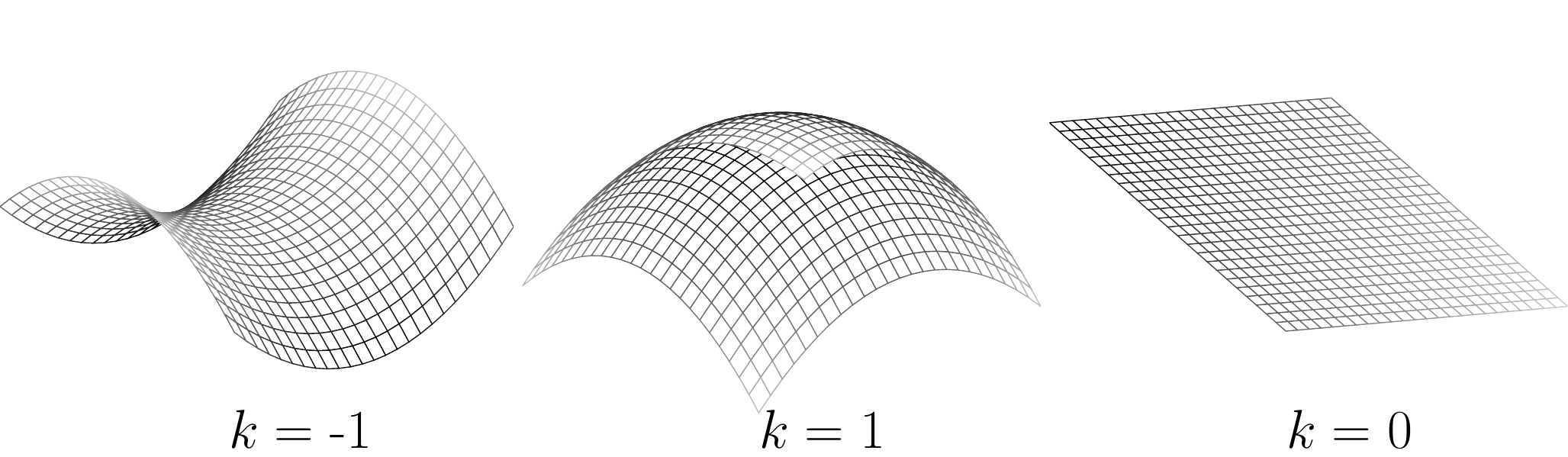}
\caption{Representação das três possibilidades de curvaturas: $k=-1$, $k=1$ e $k=0$, respectivamente.}
\label{geometrias}
\end{minipage}
\end{figure}
Podemos considerar a métrica que descreve um Universo homogêneo e isotrópico, conhecido como métrica de Robertson-Walker~\cite{robertson1,robertson2,robertson3,walker}, cujo elemento de linha é dado por
\begin{widetext}
\begin{equation}
\label{rw}
ds^{2}=c^{2}dt^{2}-a(t)^{2}\left(\frac{dr^{2}}{1-kr^{2}/R_{0}^{2}}+r^{2}d\theta^{2}+r^{2}\sin^{2}\theta d\phi^{2}\right)\;,
\end{equation}
\end{widetext}
onde $k$ é o termo de curvatura, que nos diz a geometria do Universo e pode assumir os valores $k=1$ para  uma geometria esférica, $k=-1$ para a geometria hiperbólica e $k=0$ para a geometria plana. Tais possibilidades podem ser visualizadas em uma versão bidimensional na \autoref{geometrias}. Temos ainda que, $R_{0}$ é o raio de curvatura do Universo. Por fim, $a(t)$ é o fator de escala que, como o próprio nome diz, define a escala das distâncias em um Universo homogêneo e isotrópico, tal que $a(t_{0})=1$ no presente, onde $t=t_{0}$~\cite{barbara}. 

Visto que a Eq.~(\ref{rw}) é o elemento de linha que descreve a dinâmica de um Universo em concordância com o Princípio Cosmológico, temos que $r$, $\theta$ e $\phi$ são as \textit{comoving coordinates}. Sendo assim, o tensor métrico, $g_{\mu\nu}$, descreve um Universo de acordo com Princípio Cosmológico, já que $ds^{2}=g_{\mu\nu}dx^{\mu}dx^{\nu}$. Esta mesma simetria deve estar presente no tensor energia-momento e, como consequência, no escalar energia-momento. A escolha mais simples que podemos imaginar é aquela referente ao tensor, $T_{\mu\nu}$, de um fluido perfeito, i.e., um fluido que é caracterizado por uma densidade $\rho$, uma quadri-velocidade $U_{\mu}$ e uma pressão, $P$, atuando igualmente em todas as direções. {Além disso, devido a homogeneidade e isotropia, $\rho$ e $P$ dependem somente do tempo.} Temos, dessa forma
\begin{equation}
\label{tensorfluidoperfeito}
\left[T_{\mu\nu}\right]=\left(\begin{array}{cccc}

                              \rho c^{2} & 0 & 0 & 0\\

                              0          & -P & 0 & 0\\

                              0          & 0 & -P & 0\\

                              0          & 0 & 0 & -P

\end{array}\right)\;.
\end{equation}
{No Apêndice B do material suplementar demonstramos que a partir da conservação do tensor energia-momento pode se derivar a  \textbf{equação de fluido} dada por}
\begin{equation}
\label{equaçao_fluido}
\dot{\rho}+\frac{3}{c^{2}}\frac{\dot{a}}{a}\left(\rho c^{2}+P\right)=0\;.
\end{equation}

Com os coeficientes métricos que podemos extrair da Eq.~(\ref{rw}), utilizando que $ds^{2}=g_{\mu\nu}dx^{\mu}dx^{\nu}$, e o tensor energia-momento dado pela Eq.~(\ref{tensorfluidoperfeito}), podemos resolver a Eq.~(\ref{eqdecampo}) e obter a \textbf{equação de aceleração}
\begin{equation}
\label{eq_aceleraçao_lambda}
\frac{\ddot{a}}{a}=-\frac{4\pi G}{3c^{2}}\left(\rho c^{2}+3P\right)+\frac{\Lambda}{3}\;,
\end{equation}
e a \textbf{equação de Friedmann}~\cite{friedmann1922,friedmann1924}
\begin{equation}
\label{friedmann}
\left(\frac{\dot{a}}{a}\right)^{2}=\frac{8\pi G}{3}\rho-\frac{kc^{2}}{R_{0}^{2}a^{2}}+\frac{\Lambda}{3}\;,
\end{equation}
cujas derivações se encontram no Apêndice~A do material suplementar. Salientamos que estamos utilizando a notação de Einstein para as derivadas temporais, i.e., com um ponto sobre a variável. 

As equações (\ref{equaçao_fluido})-(\ref{friedmann}) não são linearmente independentes, i.e., podemos obter a Eq.~(\ref{eq_aceleraçao_lambda}) combinando as equações (\ref{equaçao_fluido}) e (\ref{friedmann}). Assim, temos duas equações e três incógnitas: $\rho(t)$, $P(t)$ e $a(t)$. Precisamos, então, de uma quarta equação para fechar o conjunto mínimo de equações necessárias para descrevermos a evolução do Universo. Para tanto, utilizamos uma equação de estado, que relaciona a pressão e a densidade dos constituintes do Universo. {Definir a equação de estado significa especificar a natureza dos constituintes do Universo, estabelecendo assim uma conexão direta entre Cosmologia e Física de Partículas}. Para os nossos propósitos, iremos considerar a seguinte \textbf{equação de estado}~\cite{cheng2}
\begin{equation}
\label{eq_estado}
P=\omega c^{2}\rho\;,
\end{equation}
onde $\omega$ é um parâmetro adimensional que caracteriza um dado constituinte com densidade $\rho$. Podemos obter a evolução da densidade de uma dada componente, inserindo a Eq.~(\ref{eq_estado}) na Eq.~(\ref{equaçao_fluido}), o que resulta
\begin{equation}
\label{evoluçao_densidade}
\rho_{\omega}(a)=\rho_{\omega,0}a^{-3(1+\omega)}\;,
\end{equation}
onde $\rho_{\omega,0}$ é a densidade da componente $\omega$ no presente.

A partir da equação de Friedmann podemos derivar a densidade crítica do Universe, a qual é a densidade limiar entre as possíveis geometrias 
\begin{equation}
\label{densidade_critica}
\rho_{c}(t)\equiv\frac{3H(t)^{2}}{8\pi G}\;,
\end{equation}
sendo que  $H(t)=\dot{a}/a$ é o parâmetro de Hubble tal que, se $t=t_{0}$, temos $H(t_{0})=H_{0}$, a chamada constante de Hubble {definida como~\cite{lars,pdg2020}
\begin{equation}
\label{const_H}
H_{0}=100\,h~\textrm{km~s}^{-1}\textrm{Mpc}^{-1}\;,
\end{equation}
onde $h$ é  o parâmetro de Hubble escalonado, cujo valor é $h=0,674\pm0,005$~\cite{pdg2020}. Consequentemente,  o valor da constante de Hubble é ~\cite{Aghanim:2018eyx,pdg2020}
\begin{equation}
\label{const_hubble}
H_{0}=(67,4\pm0,5)\textrm{km~s}^{-1}\textrm{Mpc}^{-1}\;.
\end{equation}
Este valor, obtido pela Colaboração Planck a partir das anisotropias da radiação cósmica de fundo \cite{Aghanim:2018eyx},  será utilizado nos demais cálculos apresentados neste artigo.  Entretanto, é importante enfatizar que o estudo realizado na Ref. 
\cite{Riess:2019cxk}, que deriva $H_0$ a partir de medidas de distância de galáxias no Universo local, usando variáveis cefeidas e supernovas tipo Ia (SNe Ia), resulta em $H_0 = 74,03 \pm 1,42 \textrm{km~s}^{-1}\textrm{Mpc}^{-1}$. Esta discrepância em mais de 4$\sigma$ de confiança estatística é um importante problema em aberto no modelo padrão da Cosmologia, sendo usualmente denominado {\it $H_0$ -- tension problem}. Para uma discussão mais detalhada, recomendamos as Refs. 
\cite{Freedman:2017yms,Verde:2019ivm} .}

{Inserindo a  Eq.~(\ref{const_H}) na Eq.~(\ref{densidade_critica}) resulta que o valor da densidade crítica, no presente, é de~\cite{pdg2020}
\begin{equation}
\label{densidade_critica_atual}
\rho_{c,0}=1,878\times10^{-26}h^{2}~\textrm{kg~m}^{-3}\;.
\end{equation}
}Se a densidade do Universo for menor do que a densidade crítica dada pela Eq. (\ref{densidade_critica_atual}) a geometria do Universo será hiperbólica, enquanto que, se a densidade for maior do que a densidade crítica, a geometria será esférica e, por fim, se a densidade for igual a densidade crítica, a geometria será plana. 

{As equações de Friedmann, de aceleração, de fluido e de estado são equações necessárias para descrevermos a evolução do Universo. A fim de resolvê-las devemos especificar os constituintes presentes no Universo, o que será feito  a seguir.}
\subsection{A Composição}
\label{composição}
A componente mais notória que podemos considerar é a matéria ordinária que integra as estruturas observáveis (e.g.~planetas, estrelas, galáxias). Dado que a densidade total do universo é extremamente baixa ($\sim 10^{-27}\textrm{kg~m}^{-3}$), podemos considerar um gás que é governado pela equação dos gases ideais e obter~\cite{callen}
\begin{equation}
\label{gases_ideais}
P=\frac{\langle v^{2}\rangle}{3}\rho\;,
\end{equation}
onde $\langle v^{2}\rangle$ é a velocidade quadrática média das partículas que compõem o gás. Se compararmos a Eq.~(\ref{gases_ideais}) com a Eq.~(\ref{eq_estado}), resulta
\begin{equation}
\label{omega}
\omega=\frac{\langle v^{2}\rangle}{3c^{2}}\;.
\end{equation}
Como a matéria ordinária é composta por partículas não relativísticas, decorre que $\langle v^{2}\rangle\ll c^{2}$, o que implica $\langle v^{2}\rangle/3c^{2}\approx 0$.  Consequentemente, da Eq. (\ref{omega}), resulta
\begin{equation}
\label{omega_materia}
\omega_{m}\approx0\;.
\end{equation}
Inserindo a Eq.~(\ref{omega_materia}) na Eq.~(\ref{evoluçao_densidade}), obtemos que a densidade de matéria evolui com~\cite{weinberg,cheng2} 
\begin{equation}
\label{evoluçao_densidade_materia}
\rho_{m}(a)=\rho_{m,0}a^{-3}\;.
\end{equation}
Esse é um resultado esperado pois para uma distribuição esférica de matéria tem-se que conforme o volume da distribuição aumenta, a densidade de matéria diminui com o raio ao cubo. Como o raio é diretamente proporcional ao fator de escala, resulta que a densidade deve decrescer com o fator de escala ao cubo.

Em meados dos anos 30, o astrônomo Fritz Zwicky (1898--1974) propôs a existência de uma nova forma de matéria, distinta da matéria bariônica, denominada matéria escura~\cite{zwicky}. Enquanto observava o aglomerado de Coma, Zwicky notou que a soma das contribuições de massa das estrelas e gases nesse aglomerado não era suficiente para que a atração gravitacional mantivesse as galáxias unidas, chegando a conclusão de que deveria haver mais massa do que o observado~\cite{zwicky}. Além disso, essa matéria não deveria interagir eletromagneticamente devido ao fato de não ser possível observá-la\footnote{Por isso é chamada de matéria escura.} e, portanto, deveria ser de natureza distinta à da matéria ordinária~\cite{froes}.

Outra evidência acerca da existência de matéria escura diz respeito ao famoso problema da curva de rotação das galáxias. Vera Rubin (1928--2016) e colaboradores efetuaram curvas de rotações a partir de observações, onde conseguiram extrapolar tais observações para além das regiões visíveis das galáxias~\cite{vera1980}. Verificou-se que, ao invés de a velocidade tangencial de um dado observável diminuir com $r^{-1/2}$ a partir da região visível da galáxia, ela permanece constante. {Como consequência, mesmo após à região visível espera-se a presença de matéria}, mas que não pode ser observada diretamente, retomando a proposta da existência de matéria escura {sugerida} por Zwicky~\cite{cheng}. Não obstante, para efeitos de melhor ajuste com os dados observacionais, a matéria escura precisa ser não relativística, o que justifica a nomenclatura matéria escura fria, proveniente do inglês \textit{cold dark matter} (CDM)~\cite{froes}.

Sendo assim, levando-se em conta a presença de matéria escura, e sendo ela de natureza atrativa e não relativística, assim como a matéria ordinária, quando consideramos a Eq.~(\ref{evoluçao_densidade_materia}), estamos considerando tanto matéria ordinária quanto matéria escura. Logo, daqui por diante, quando nos referirmos à componente matéria estamos nos referindo à matéria ordinária e à matéria escura. Salientamos que a natureza da matéria escura ainda é desconhecida e a busca para determiná-la diz respeito à uma extensa e rica área da Física~\cite{dark_matter,dark_review}.

Outra componente notória é a radiação, onde englobamos nela não somente fótons, mas também quaisquer outras partículas relativísticas. Dessa forma, teremos que $\langle v^{2}\rangle/c^{2} \approx 1$ onde, com isso, a Eq.~(\ref{omega}) nos diz que 
\begin{equation}
\label{omega_radiaçao}
\omega_{r}\approx\frac{1}{3}\;.
\end{equation}
Se inserirmos este resultado na Eq.~(\ref{evoluçao_densidade}), resulta que a densidade de radiação evolui com~\cite{weinberg,cheng2}
\begin{equation}
\label{evoluçao_densidade_radiaçao}
\rho_{r}(a)=\rho_{r,0}a^{-4}\;.
\end{equation}
Esse resultado também é esperado. Primeiramente, temos que devido ao acréscimo do volume, a densidade de radiação irá variar com $\propto a^{-3}$. O fator extra é devido à variação do comprimento de onda com a expansão. Dado que o comprimento de onda $\lambda$ é proporcional a $a${, bem como} a energia é dada por $E=hc/\lambda$, tem-se que ela é proporcional a $a^{-1}$. Consequentemente, temos que a densidade de radiação varia com $\propto a^{-4}$~\cite{liddle}.

Como apontado anteriormente, se considerarmos apenas matéria e radiação não obtemos {a observada expansão acelerada do Universo}~\cite{perlmutter,perlmutter2,riess} e, dessa forma, precisamos inserir a Constante Cosmológica. Podemos obter $\omega_{\Lambda}$ da seguinte maneira. Perceba que, se na Eq.~(\ref{friedmann}) nós suprimirmos o termo referente à Constante Cosmológica e fizermos $\rho\rightarrow\rho+\rho_{\Lambda}$, podemos recuperar a Eq.~(\ref{friedmann}) com o fator $\Lambda/3$, se 
\begin{equation}
\label{densidade_lambda}
\rho_{\Lambda}=\frac{\Lambda}{8\pi G}\;\,,
\end{equation}
ou seja, a densidade associada à constante cosmológica é constante durante a evolução temporal do Universo.
A partir de $\rho_{\Lambda}$ podemos derivar a pressão associada ao termo cosmológico {inserindo a Eq.~(\ref{densidade_lambda}) na Eq.~(\ref{equaçao_fluido}), o qual resulta}
\begin{equation}
P_{\Lambda}=-\frac{\Lambda}{8\pi G}c^{2}=-\rho_{\Lambda}c^{2}\;,
\end{equation}
onde, por uma simples comparação com a Eq.~(\ref{eq_estado}), segue que
\begin{equation}
\label{omega_lambda}
\omega_{\Lambda}=-1\;.
\end{equation}
Veja que obtivemos uma pressão negativa para a componente associada à Constante Cosmológica, o que é esperado  pois uma pressão negativa nos proporciona a repulsão gravitacional desejada para que a expansão do Universo seja acelerada~\cite{liddle}. Assim como a matéria escura, a natureza da Constante Cosmológica ainda é uma questão em aberto~\cite{dark_review}.

\begin{figure}[t]
\centering
\begin{minipage}{8.5cm}
\centering
\includegraphics[width=8.5cm]{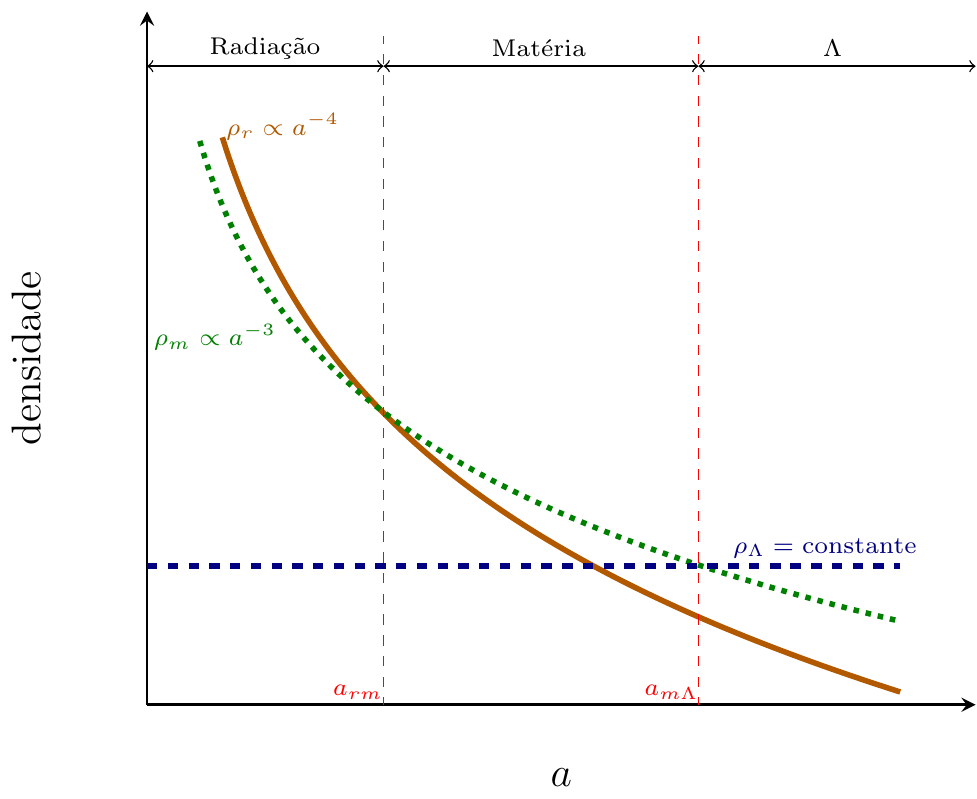}
\caption{Representação da evolução das densidades associadas à radiação (curva contínua), matéria (curva pontilhada) e Constante Cosmológica (curva tracejada). {As grandezas $a_{rm}$ e $a_{m\Lambda}$ representam fatores de escala onde há equivalência das constribuições de matéria e radiação, bem como de matéria e da constante cosmológica, respectivamente. Tais fatores de escala serão discutidos em detalhe na Seção \ref{duas_componentes}.}}
\label{evolucao_rho}
\end{minipage}
\end{figure}

Podemos visualizar qualitativamente a evolução das densidades para cada uma das componentes na \autoref{evolucao_rho}, a qual indica que a radiação deve ser dominante nos instantes iniciais ($t \rightarrow 0$), enquanto que o termo cosmológico deve determinar o seu destino final. Entre estes dois limites assintóticos, devemos ter um regime no qual a matéria é dominante.

\subsection{O Parâmetro de Densidade}
\label{parametro}
É conveniente reescrevermos a equação de Friedmann em termos de um parâmetro adimensional chamado \textbf{parâmetro de densidade}, definido como~\cite{cheng2}
\begin{equation}
\label{def_parametro_densidade}
\Omega(t)\equiv\frac{\rho(t)}{\rho_{c}(t)}\;,
\end{equation}
onde, em $t=t_{0}$, para um dado constituinte definido por $\omega$, temos que
\begin{equation}
\label{par_atual_omega}
\Omega_{\omega,0}=\frac{\rho_{\omega,0}}{\rho_{c,0}}\;.
\end{equation}
O parâmetro de densidade total, em $t=t_{0}$, será dado por
\begin{equation}
\label{par_omega}
\Omega_{0}=\frac{\rho_{0}}{\rho_{c,0}}=\sum_{\omega}\Omega_{\omega,0}=\Omega_{r,0}+\Omega_{m,0}+\Omega_{\Lambda,0}\;,
\end{equation}
onde $\rho_{0}=\rho_{r,0}+\rho_{m,0}+\rho_{\Lambda,0}$. Se lembrarmos que $\rho_{c}$ é a densidade limiar entre as possíveis geometrias, podemos perceber que o Universo terá geometria hiperbólica ($k=-1$) se $\Omega_{0}<1$, geometria esférica ($k=+1$) se $\Omega_{0}>1$ e geometria plana ($k=0$) se $\Omega_{0}=1$. 

Sendo assim, podemos reescrever a Eq.~(\ref{friedmann}) fazendo uso das equações (\ref{def_parametro_densidade})-(\ref{par_omega}), tal que 
\begin{equation}
\label{friedmann_omega}
\left(\frac{\dot{a}}{a}\right)^{2}=H_{0}^{2}\left(\frac{\Omega_{r,0}}{a^{4}}+\frac{\Omega_{m,0}}{a^{3}}+\Omega_{\Lambda,0}-\frac{\Omega_{0}-1}{a^{2}}\right)\;.
\end{equation}
Os parâmetros de densidade são obtidos através de observações onde, atualmente, temos~{\cite{Aghanim:2018eyx,pdg2020}
\begin{eqnarray}
\label{par_valores}
\Omega_{r,0}&=&2,47\times10^{-5}~h^{-2}\;,\nonumber\\
\Omega_{m,0}&=&0,315\pm0,007\;,\nonumber\\
\Omega_{\Lambda,0}&=&0,685\pm0,007\;.
\end{eqnarray}
Uma discussão detalhada sobre o procedimento para obter os principais parâmetros cosmológicos a partir dos dados experimentais disponíveis na literatura é apresentada na Ref.~\cite{pordeus}, a qual recomendamos fortemente ao leitor interessado no tema.}

{Com os elementos apresentados nesta Seção  estamos aptos a obter as soluções da equação de Friedmann~(\ref{friedmann_omega}) através da evolução do fator de escala em função do tempo. Devido ao fato de haver uma grande quantidade de cenários possíveis através das escolhas de composição e geometria, vamos nos restringir aos cenários mais instrutivos no que se refere à construção, gradativa, do cenário atual.}
\section{A Influência da Curvatura}
\label{curvatura}
Nesta Seção analisamos como diferentes considerações referentes à geometria influenciam as soluções da equação de Friedmann. Para analisarmos tais possibilidades, consideramos um Universo composto apenas por matéria, onde $\Omega_{r,0}=0$ e $\Omega_{\Lambda,0}=0$. Portanto, decorre da Eq.~(\ref{friedmann_omega}) que~\cite{barbara,roos}
\begin{equation}
\label{mat_curv}
\left[H(t)\right]^{2} \equiv \left(\frac{\dot{a}}{a}\right)^{2}=H_{0}^{2}\left(\frac{\Omega_{0}}{a^{3}}-\frac{\Omega_{0}-1}{a^{2}}\right)\;.
\end{equation}
{Desta equação obtemos
\begin{equation}
\label{ed_materia_curvatura}
\dot{a}=H_{0}\left(\frac{\Omega_{0}}{a}-\Omega_{0}+1\right)^{1/2}\;,
\end{equation}
a qual pode ser resolvida por integração direta, tal que}
\begin{equation}
\label{resol_ed_materia_curvatura}
\int_{0}^{a}d\tilde{a}\left(\frac{\Omega_{0}}{\tilde{a}}-\Omega_{0}+1\right)^{-1/2}=H_{0}t\;.
\end{equation}
{No que segue iremos analisar a influência da curvatura através da escolha de distintos valores para $\Omega_{0}$.}
\subsection{Geometria esférica}
\label{materia_esferico}
Consideremos um Universo atualmente em expansão, logo, $H_{0}>0$. Como há somente matéria, através da Lei da Gravitação de Newton podemos esperar que tal expansão em algum momento cesse, iniciando posteriormente um período de contração. No momento de estagnação da expansão devemos ter $H(t)=0$. A Eq.~(\ref{mat_curv}) nos diz que $H(t)=0$, se~\cite{barbara,roos,kolb}
\begin{equation}
\label{maximo}
\frac{\Omega_{0}}{a_{\textrm{\tiny{máx}}}^{3}}-\frac{\Omega_{0}-1}{a_{\textrm{\tiny{máx}}}^{2}}=0\quad\Rightarrow\quad a_{\textrm{\tiny{máx}}}=\frac{\Omega_{0}}{\Omega_{0}-1}\;,
\end{equation}
onde $a_{\textrm{\tiny{máx}}}$ é o fator de escala no qual há a estagnação da expansão. Além disso, como $\Omega_{0}>0$ teremos $\Omega_{0}/a^{3}>0$ e, portanto, para que possamos obter $H(t)=0$, é necessário que~\cite{barbara}
\begin{equation}
-\frac{\Omega_{0}-1}{a^{2}}<0\quad\Rightarrow\quad\Omega_{0}>1\;.
\end{equation}
Sendo assim, considerando um Universo cuja única componente seja matéria, para que haja um período de contração é necessário que a densidade atual seja maior do que a densidade crítica, o que implica que devemos ter um Universo com geometria esférica~\cite{barbara,liddle}.

A solução da Eq.~(\ref{resol_ed_materia_curvatura}) para $\Omega_{0}>1$, i.e., para $k=+1$, 
{pode ser obtida através da introdução do chamado ângulo de desenvolvimento $\alpha$~\cite{weinberg2,kolb}, tal que
\begin{equation}
\label{develop_angle}
1-\cos{\alpha} = \frac{\Omega_{0}-1}{\Omega_{0}}2a\;.
\end{equation}
Dessa forma, se fizermos uma substituição do tipo~\cite{narlikar_intro}
\begin{equation}
\label{subst}
\tilde{a}=\frac{\Omega_{0}}{\Omega_{0}-1}\sin^{2}{\left(\frac{\alpha}{2}\right)}=\frac{\Omega_{0}}{\Omega_{0}-1}\left(\frac{1-\cos{\alpha}}{2}\right)\;,
\end{equation}
obtemos a seguinte solução parametrizada para a Eq.~(\ref{resol_ed_materia_curvatura})~\cite{barbara}
\begin{equation}
\label{a_materia_curvatura1}
a(\alpha)=\frac{\Omega_{0}}{\Omega_{0}-1}\left(\frac{1-\cos\alpha}{2}\right)\;,
\end{equation}
e
\begin{equation}
\label{t_materia_curvatura1}
t(\alpha)=\frac{1}{H_{0}}\frac{\Omega_{0}}{(\Omega_{0}-1)^{3/2}}\left(\frac{\alpha-\sin\alpha}{2}\right)\;.
\end{equation}
Note que $\alpha$ deve ser definido em $0\leq\alpha\leq2\pi$ de forma que, a expansão atinge o seu máximo em $\alpha = \pi$, iniciando, posteriormente, o período de contração até encerrar em uma grande implosão, denominada \textit{Big Crunch}, quando $\alpha = 2\pi$~\cite{weinberg2,barbara,narlikar_intro}.} A representação gráfica do fator de escala em função do tempo, dada pelas equações (\ref{a_materia_curvatura1}) e (\ref{t_materia_curvatura1}), pode ser visualizada na \autoref{graf_mat_curv} através da curva pontilhada.

Por meio da solução dada pelas equações (\ref{a_materia_curvatura1}) e (\ref{t_materia_curvatura1}) podemos notar algumas características interessantes desse cenário. Primeiramente, a Eq.~(\ref{a_materia_curvatura1}) nos mostra que obteremos o fator de escala máximo quando $\alpha=\pi$ e, com isso, confirmamos o que já demonstramos anteriormente na Eq.~(\ref{maximo}). Em segundo lugar, por meio da Eq.~(\ref{t_materia_curvatura1}) podemos extrair o "tempo máximo de vida", $t_{\textit{\tiny{crunch}}}$, de um Universo dotado de geometria esférica e composto apenas por matéria. É facil ver que isso ocorre quando $\alpha=2\pi$, logo~\cite{barbara,roos}
\begin{equation}
\label{t_crunch}
t_{\textit{\tiny{crunch}}}=\frac{\pi}{H_{0}}\frac{\Omega_{0}}{(\Omega_{0}-1)^{3/2}}\;.
\end{equation}
Dizemos, portanto, que um Universo positivamente curvado e composto apenas por matéria irá se expandir até o fator de escala máximo dado pela Eq.~(\ref{maximo}) e, então, irá colapsar no que chamamos de \textit{Big Crunch}~\cite{barbara,roos,das}.
\subsection{Geometria hiperbólica}
\label{materia_hiperbolico}
A solução da Eq.~(\ref{resol_ed_materia_curvatura}) para $\Omega_{0}<1$ pode ser 
{obtida seguindo a mesma lógica do caso anterior, com a diferença de que agora o ângulo de desenvolvimento é imaginário, i.e., $\alpha = i\beta$~\cite{kolb,weinberg2}. Dessa forma, teremos
\begin{equation}
\label{develop_angle2}
1-\cosh{\beta}=\frac{\Omega_{0}-1}{\Omega_{0}}2a\;.
\end{equation}
Dito isso, a substituição necessária para resolver a Eq.~(\ref{resol_ed_materia_curvatura}) é dada por~\cite{weinberg2}
\begin{equation}
\label{subst2}
\tilde{a}=-\frac{\Omega_{0}}{\Omega_{0}-1}\sinh^{2}{\left(\frac{\beta}{2}\right)}=-\frac{\Omega_{0}}{\Omega_{0}-1}\left(\frac{\cosh{\beta}-1}{2}\right)\;,
\end{equation}
com a qual obtemos a solução parametrizada~\cite{barbara}
\begin{equation}
\label{a_materia_curvatura2}
a(\beta)=\frac{\Omega_{0}}{1-\Omega_{0}}\left(\frac{\cosh\beta-1}{2}\right)\;,
\end{equation}
e
\begin{equation}
\label{t_materia_curvatura2}
t(\beta)=\frac{1}{H_{0}}\frac{\Omega_{0}}{(1-\Omega_{0})^{3/2}}\left(\frac{\sinh\beta-\beta}{2}\right)\;.
\end{equation}
Nesse caso, temos que $0\leq\beta<\infty$~\cite{kolb,weinberg2}.} Podemos visualizar graficamente as equações (\ref{a_materia_curvatura2}) e (\ref{t_materia_curvatura2}) através da curva tracejada na \autoref{graf_mat_curv}.

Tendo em vista a solução da Eq.~(\ref{resol_ed_materia_curvatura}) para um Universo negativamente curvado e composto apenas por matéria, dado pelas equações (\ref{a_materia_curvatura2}) e (\ref{t_materia_curvatura2}), podemos perceber que, em tal cenário, considerando que atualmente o Universo está em expansão, teremos que esse irá se expandir indefinidamente ao que chamamos de \textit{Big Chill}~\cite{barbara,das}.
\subsection{Geometria plana}
\label{materia_plano}
\begin{figure}[t]
\centering
\begin{minipage}{8.5cm}
\centering
\includegraphics[width=8.5cm]{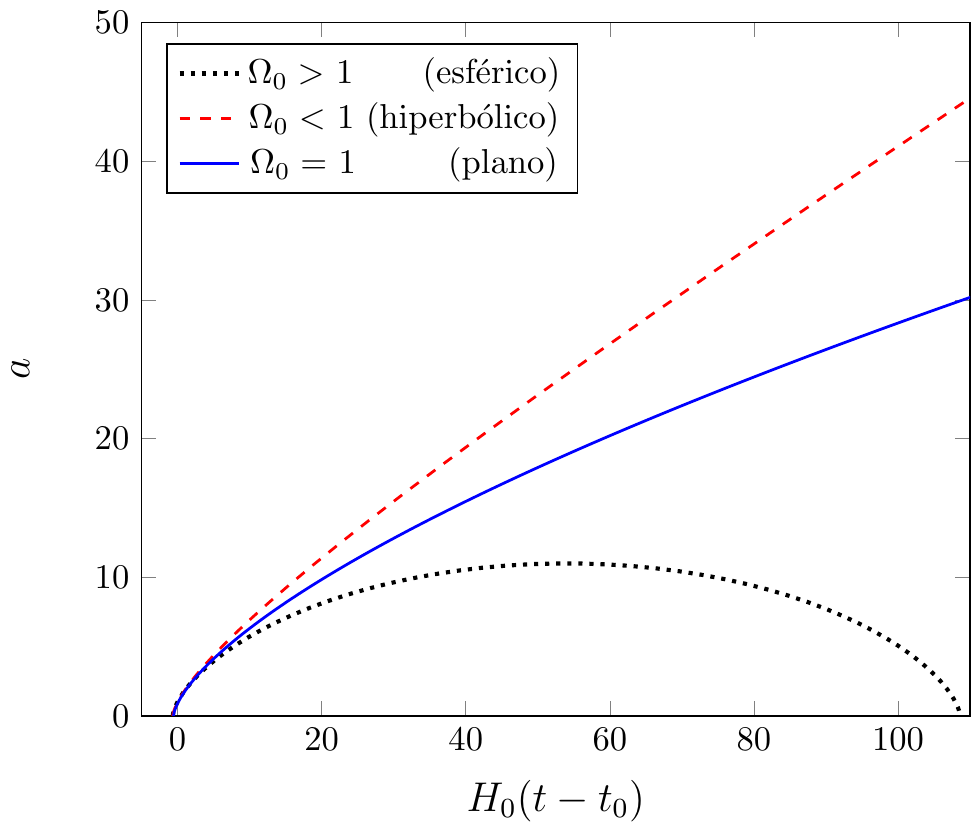}
\caption{Fator de escala em função do tempo para um Universo dominado por matéria espacialmente plano (curva contínua), esférico (curva pontilhada) e hiperbólico (curva tracejada).}
\label{graf_mat_curv}
\end{minipage}
\end{figure}

Finalizando as possibilidades de geometria para um Universo composto apenas por matéria, temos o Universo onde $\Omega_{0}=1$, ou seja, um Universo onde a densidade é comparável à densidade crítica, corriqueiramente chamado de Universo de Einstein - de Sitter~\cite{roos,lars}.  Para esse cenário, podemos reescrever a Eq.~(\ref{resol_ed_materia_curvatura}) como
\begin{equation}
\label{ed_materia_plano}
\int_{0}^{a}\tilde{a}^{1/2}d\tilde{a}=H_{0}t\;.
\end{equation}
Resolvendo a Eq.~(\ref{ed_materia_plano}), resulta~\cite{barbara,liddle,lars}
\begin{equation}
\label{a_materia_plano}
a(t)=\left(\frac{t}{t_{0}}\right)^{2/3}\;,
\end{equation}
onde~\cite{liddle,weinberg,cheng}
\begin{equation}
\label{idade_materia}
t_{0}=\frac{2}{3H_{0}}\;,
\end{equation}
{é a idade do Universo para o cenário em questão. Para completeza do nosso estudo,  os valores de $t_0$ para as geometrias esférica e hiperbólica são derivados no Apêndice C.  A solução dada pela Eq.~(\ref{a_materia_plano}) é representada pela curva contínua na \autoref{graf_mat_curv}.}
\subsection{Análise Comparativa}
Perceba da \autoref{graf_mat_curv} que, para tempos pequenos, as três possibilidades de curvatura demonstram um comportamento do fator de escala extremamente similares. Para o caso onde $k=0$ ($\Omega_{0}=1$), durante toda a expansão teremos $a\propto t^{2/3}$. Já nos casos onde $k\neq0$ ($\Omega_{0}>1$ e $\Omega_{0}<1$) decorre que, na Eq.~(\ref{mat_curv}), $\Omega_{0}a^{-3}\gg(1-\Omega_{0})a^{-2}$, o que implica que podemos aproximar, no limite de $t$ pequeno, pelo comportamento do fator de escala do cenário para geometria plana\footnote{Uma análise minuciosa a respeito das soluções clássicas da equação de Friedmann efetuadas na Seção~\ref{curvatura} no que diz respeito ao intervalo de pequenos valores de tempo á apresentada na referência~\cite{viglioni}.}.

A partir de $t\sim H_{0}^{-1}$, conhecido como tempo de Hubble, podemos notar que se inicia uma diferenciação mais acentuada entre as três curvas. Isto ocorre porque o conteúdo de matéria se dilui devido a expansão e, consequentemente, $\Omega_{0}a^{-3}\ll(1-\Omega_{0})a^{-2}$. Com isso, temos que, um Universo espacialmente plano continuará expandindo-se com $a\propto t^{2/3}$. Porém, um Universo espacialmente hiperbólico irá se expandir similarmente ao Universo de Milne, i.e., um Universo com $k=-1$ ($\Omega_{0}<1$) e vazio~\cite{lars}. Isso é o que chamamos de expansão livre~\cite{liddle}. O caso onde $k=+1$ ($\Omega_{0}>1$) possui um destino diferente aos dois anteriores, qual seja, não irá findar em um \textit{Big Chill}. Nesse caso o conteúdo de matéria não se diluirá o suficiente devido ao fato de que, sendo a sua densidade maior do que a densidade crítica, em uma interpretação newtoniana, a quantidade de matéria permite que a atração gravitacional sobreponha a expansão e inicie uma contração~\cite{liddle}. Sendo assim, nesse cenário haverá uma expansão até o fator de escala máximo dado pela Eq.~(\ref{maximo}) e, então, inicia-se o período de contração até que, enfim, colapse em um \textit{Big Crunch}~\cite{cheng,barbara}. Note que o período de contração é simétrico ao período de expansão, i.e., podemos fazer uma substituição de $t$ por $-t$ na Eq.~(\ref{mat_curv}) e mesmo assim ela permanecerá a mesma~\cite{liddle}. Isso só ocorre devido ao fato de que estamos tratando de um Universo homogêneo e isotrópico onde, dessa forma, a expansão é adiabática~\cite{roos}, nos permitindo considerar a evolução do Universo como um processo reversível\footnote{Se o Universo é homogêneo e isotrópico, a temperatura será a mesma em todo o Universo e, consequentemente, não há fluxo de calor em uma certa região de volume $dV$. Portanto, da Termodinâmica, temos que $dQ=TdS=0$. Logo, a entropia não varia e podemos considerar o processo como sendo reversível.}~\cite{barbara}. Deste modo, o parâmetro de densidade além de determinar a geometria do espaço, também nos diz qual será o destino do Universo~\cite{cheng2}. Podemos observar tais possibilidades na \autoref{destino}. 

Dados observacionais indicam que o Universo é espacialmente plano~\cite{WMAP_1,pdg2020}. Em vista disso, no que se segue, iremos assumir $k=0$ e focaremos em  obter as soluções da equação de Friedmann para diferentes possibilidades de composição do Universo. Primeiramente iremos relembrar o caso onde a única componente é a matéria, seguindo para o cenário onde há apenas radiação e findando com um Universo espacialmente plano composto apenas pela componente associada à Constante Cosmológica.
    \begin{table*}[t]
    \centering
        \caption{Possibilidades de comportamento do Universo para tempos maiores do que $H_{0}^{-1}$.}
        \label{destino}
        \begin{tabular}{ccc}
            \hline
            Parâmetro de Densidade\;\;\;\;\;\;\;\;\;&Termo de Curvatura\;\;\;\;\;\;\;\;\;&Comportamento para $t\gg H_{0}^{-1}$\\
            \hline
            \hline
            $\Omega_{0}<1$&$k=-1$&\textit{Big Chill} ($a\propto t$)\\
            $\Omega_{0}=1$&$k=0$&\textit{Big Chill} ($a\propto t^{2/3}$)\\
            $\Omega_{0}>1$&$k=+1$&\textit{Big Crunch}\\
            \hline
        \end{tabular}\\
    Adaptado de~\cite{barbara}
    \end{table*}
\section{Universo espacialmente plano composto por apenas uma componente}
\label{single_component}
\begin{figure}[t]
\centering
\begin{minipage}{8.5cm}
\centering
\includegraphics[width=8.5cm]{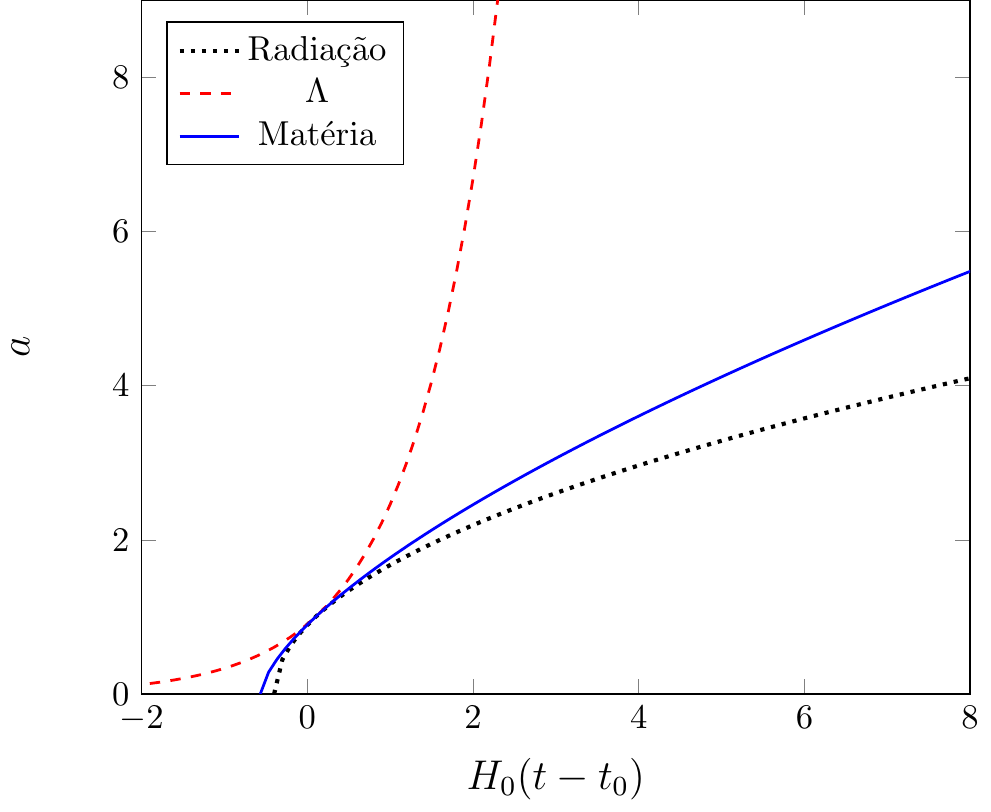}
\caption{Fator de escala em função do tempo para um Universo espacialmente plano e composto apenas por matéria (curva contínua), radiação (curva pontilhada) e $\Lambda$ (curva tracejada).}
\label{graf_mat_rad_lambda}
\end{minipage}
\end{figure}
\subsection{Universo Composto por Matéria}
\label{materia}
Como visto na Seção~\ref{materia_plano}, temos que um Universo dotado de geometria plana e composto apenas por matéria se expande com um fator de escala dado por~\cite{barbara,liddle,lars}
\begin{equation}
\label{a_materia_plano2}
a(t)=\left(\frac{t}{t_{0}}\right)^{2/3}\;.
\end{equation}
Além disso, um Universo com essas características possui uma idade de~\cite{liddle,weinberg,cheng}
\begin{equation}
\label{idade_materia2}
t_{0}=\frac{2}{3H_{0}}\;,
\end{equation}
{onde, utilizando o valor da constante de Hubble dado pela pela Eq.~(\ref{const_hubble}), temos que essa idade é de
\begin{equation}
\label{idade_materia3}
t_{0}=9,68\times10^{9}\textrm{\,anos}\;. 
\end{equation} 
É importante enfatizar que este valor é inferior àquele obtido para a idade das estrelas mais antigas presentes em globulares estelares, as quais tem uma idade superior a 10$^{9}$ anos. Portanto, um Universo composto apenas por matéria não é um cenário viável para descrever nosso Universo.}
Representamos a solução dada pela Eq.~(\ref{a_materia_plano2}) através da curva contínua na \autoref{graf_mat_rad_lambda}. 
\subsection{Universo Composto por Radiação}
\label{radiaçao}
A análise de um Universo composto por radiação é de extrema importância dado que, para tempos primordiais, o termo referente à radiação na Eq.~(\ref{friedmann_omega}) domina sobre todos os outros termos~\cite{narlikar_intro}. Para tanto, consideramos apenas a componente referente a radiação na Eq.~(\ref{friedmann_omega}) bem como $k=0$, obtendo
\begin{equation}
\label{ed_radiaçao}
\dot{a}=\frac{H_{0}}{a}\;,
\end{equation}
onde estamos levando em consideração que $\Omega_{r,0}=\Omega_{0}=1$, devido ao fato de estarmos lidando com um Universo espacialmente plano~\cite{barbara}. A solução da equação diferencial (\ref{ed_radiaçao}), obtida por integração direta, é~\cite{barbara,liddle}
\begin{equation}
\label{a_radiaçao}
a(t)=\left(\frac{t}{t_{0}}\right)^{1/2}\;,
\end{equation}
onde~\cite{barbara,cheng}
\begin{equation}
\label{idade_radiaçao}
t_{0}=\frac{1}{2H_{0}}\;,
\end{equation}
é a idade do Universo para o presente cenário {que, ao fazermos uso da Eq.~(\ref{const_hubble}), podemos obter 
\begin{equation}
\label{idade_radiaçao2}
t_{0}= 7,26\times10^{9}\textrm{\,anos}\;.
\end{equation}
Assim como um Universo composto apenas por matéria, este cenário é incompatível com os dados para as idades das estrelas mais antigas observadas}.  
A representação gráfica da solução Eq.~(\ref{a_radiaçao}) pode ser visualizada através da curva pontilhada na \autoref{graf_mat_rad_lambda}.

Perceba que, um Universo cuja única componente é a radiação e que seja dotado de uma geometria plana, expande mais lentamente se comparado a um Universo plano dominado por matéria. Isso se deve ao fato de que, na Eq.~(\ref{eq_aceleraçao_lambda}), um Universo plano e dominado por radiação, ao contrário de um Universo plano contendo apenas matéria, possui o termo referente à pressão que reduz a aceleração~\cite{liddle}. 
\subsection{Universo Composto pelo termo cosmológico}
\label{lambda}
{Finalizando a análise dos cenários compostos} por uma única componente, temos aquele onde a componente é a Constante Cosmológica, também conhecido como Universo de de Sitter~\cite{barbara}. Novamente considerando um Universo espacialmente plano, teremos que a Eq.~(\ref{friedmann_omega}) pode ser escrita como~\cite{barbara,lars}
\begin{equation}
\label{friedmann_lambda}
\frac{\dot{a}}{a}=H_{0}\;,
\end{equation}
onde, como $k=0$, $\Omega_{\Lambda,0}=\Omega_{0}=1$~\cite{barbara}. Podemos notar que $H(t)=H_{0}$, o que implica que a taxa de expansão é constante durante toda a evolução do Universo~\cite{roos}, diferentemente da taxa de expansão para um Universo plano dominado por matéria e por radiação cujas dependências temporais são de $H\propto t^{-1}$. 

Podemos resolver a Eq.~(\ref{friedmann_lambda}) por integração direta, tal que
\begin{equation}
\label{resol_friedmann_lambda}
\int_{0}^{a}\frac{d\tilde{a}}{\tilde{a}}=H_{0}\int_{t_{0}}^{t}\tilde{t}\;d\tilde{t}\;,
\end{equation}
cuja solução é dada por~\cite{barbara,roos}
\begin{equation}
\label{a_lambda}
a(t)=e^{H_{0}(t-t_{0})}\;,
\end{equation}
de modo que sua representação gráfica pode ser visualizada através da curva tracejada na \autoref{graf_mat_rad_lambda}. É importante salientar que este cenário não possui uma idade definida.

Assim, um Universo plano composto por $\Lambda$ se expande de forma exponencial~\cite{lars}. Podemos interpretar isso com o fato de que $H(t)=H_{0}$, {ou seja, não há um instante de tempo privilegiado e a razão de expansão é constante}. Outra interpretação pode ser feita lembrando-se que a pressão exercida pela Constante Cosmológica é negativa~\cite{liddle}. Assim, como a densidade $\rho_{\Lambda}$ é constante, conforme o Universo se expande maior será o seu volume e, como $dE=\rho_{\Lambda}c^{2}dV$, maior será a energia do fluido de $\Lambda$, fazendo com que o Universo se expanda ainda mais rápido~\cite{cheng2}.
\section{Universo composto por duas componentes}
\label{duas_componentes}
Nas Seções \ref{materia}, \ref{radiaçao} e \ref{lambda} obtivemos as soluções elementares para cenários onde $k=0$ e somente uma componente estava presente. Nesta Seção solucionaremos a equação de Friedmann considerando duas componentes, a qual nos aproxima um pouco mais da situação real. Consideraremos dois cenários, primeiramente matéria e radiação e, posteriormente, matéria e Constante Cosmológica.
\subsection{Universo Composto por Matéria e Radiação}
\label{mat_rad}
Considerando um Universo plano composto por matéria e radiação, temos que a razão entre as densidades de matéria e radiação é dada por
\begin{equation}
\label{razao_mat_rad}
\frac{\rho_{m}(a)}{\rho_{r}(a)}=\frac{\Omega_{m,0}}{\Omega_{r,0}}a\;,
\end{equation}
onde usamos que $\Omega_{\omega,0}=\rho_{\omega,0}/\rho_{c,0}$ e as equações (\ref{evoluçao_densidade_materia}) e (\ref{evoluçao_densidade_radiaçao}). Fazendo $\rho_{m}/\rho_{r}=1$, resulta~\cite{barbara,bari}
\begin{equation}
\label{igualdade_mat_rad}
a_{rm}=\frac{\Omega_{r,0}}{\Omega_{m,0}}\;,
\end{equation}
o qual denominamos por fator de escala de equivalência {entre radiação e matéria}. Esse fator de escala é aquele no qual radiação e matéria contribuem igualmente para com o conteúdo material do Universo. Utilizando os valores da Eq.~(\ref{par_valores}), obtemos que { 
\begin{equation}
\label{a_coexistencia_mat_rad}
a_{rm}=1,73\times10^{-4}\;.
\end{equation}}

Ademais, para o presente cenário, podemos reescrever a Eq.~(\ref{friedmann_omega}) como~\cite{barbara,hobson}
\begin{equation}
\label{ed_mat_rad}
\dot{a}=H_{0}\left(\frac{\Omega_{r,0}}{a^{2}}+\frac{\Omega_{m,0}}{a}\right)^{1/2}\;.
\end{equation}
{Isolando $\Omega_{r,0}$}  e levando em consideração a Eq.~(\ref{igualdade_mat_rad}), obtemos
\begin{equation}
\label{resol_mat_rad}
\int_{0}^{a}\frac{d\tilde{a}}{\left(1+\tilde{a}/a_{rm}\right)^{1/2}}=H_{0}\Omega_{r,0}^{1/2}t\;,
\end{equation}
que, por uma substituição do tipo $1+\tilde{a}/a_{rm}\rightarrow a'$, resulta em uma relação para $t$ em função do fator de escala, dada por~\cite{barbara}
\begin{equation}
\label{t_mat_rad}
H_{0}t=\frac{4a_{rm}^{2}}{3\Omega_{r,0}^{1/2}}\left[1+\left(1+\frac{a}{a_{rm}}\right)^{1/2}\left(\frac{a}{2a_{rm}}-1\right)\right]\;.
\end{equation}

A Eq.~(\ref{t_mat_rad}) não nos diz como será o comportamento do fator de escala em função do tempo, na forma como está apresentada. No entanto, dela podemos extrair o comportamento do fator de escala para $a\ll a_{rm}$ e $a\gg a_{rm}$, {além de determinar a idade aproximada, $t_{rm}$, quando matéria e radiação são equivalentes}.

Com a finalidade de encontrarmos $t_{rm}$, podemos simplesmente fazer $a=a_{rm}$ na Eq.~(\ref{t_mat_rad}), o que implica~\cite{barbara}
\begin{equation}
\label{t_coexistencia1}
t_{rm}=\frac{4a_{rm}^{2}}{3\Omega_{r,0}^{1/2}}\left(1-\frac{\sqrt{2}}{2}\right)H_{0}^{-1}\;,
\end{equation}
que, fazendo uso das equações (\ref{const_hubble}), (\ref{par_valores}) e (\ref{a_coexistencia_mat_rad}), resulta
{
\begin{equation}
\label{t_coexistencia2}
t_{rm}=22.951,21\textrm{\,anos}\;.
\end{equation}
}Adiante veremos que, em termos de escalas cosmológicas temporais, $t_{rm}$ é extremamente pequeno. Analisemos a solução dada pela Eq.~(\ref{t_mat_rad}) em seus limites assintóticos $a\ll a_{rm}$ e $a\gg a_{rm}$.
\subsubsection{Comportamento inicial ($a\ll a_{rm}$)}
\label{t<<t_rm}
Se considerarmos $a\ll a_{rm}$, teremos que $a/a_{rm}\ll 1$ e, consequentemente, podemos fazer a seguinte expansão
\begin{equation}
\label{exp_serie}
\left(1+\frac{a}{a_{rm}}\right)^{1/2}\approx 1+\frac{a}{2a_{rm}}-\frac{a^{2}}{8a_{rm}^{2}}\;.
\end{equation}
Se inserirmos  a Eq.~(\ref{exp_serie}) na Eq.~(\ref{t_mat_rad}) e mantivermos apenas os termos quadráticos ou de ordem inferior, resulta~\cite{barbara}
\begin{equation}
\label{t_peq_mat_rad}
a(t)\approx \left(2\Omega_{r,0}^{1/2}H_{0}t\right)^{1/2}\;.
\end{equation}

Comparando o resultado expresso na Eq.~(\ref{t_peq_mat_rad}) com o resultado expresso na Eq.~(\ref{a_radiaçao}), tem-se que para valores de tempo suficientemente pequenos, i.e., $t\ll t_{rm}$, os resultados obtidos para duas componentes serão similares aos que obtemos se considerarmos um Universo composto apenas por radiação~\cite{bari}.\footnote{Note que isso se aplica até mesmo para um Universo composto pelas três componentes e com o termo de curvatura, visto que os outros termos da Eq.~(\ref{friedmann_omega}) serão pequenos se comparados ao termos referente à radiação.}
\subsubsection{Comportamento final ($a\gg a_{rm}$)}
\label{t>>t_rm}
Prosseguindo com nossa análise assintótica da Eq.~(\ref{t_mat_rad}), temos que, para $a\gg a_{rm}$
\begin{equation}
H_{0}t\approx\frac{4a_{rm}^{2}}{3\Omega_{r,0}^{1/2}}\frac{1}{2}\left(\frac{a}{a_{rm}}\right)^{3/2}\approx\frac{2}{3}\Omega_{m,0}^{-1/2}a^{3/2}\;,
\end{equation}
onde usamos que $a_{rm}=\Omega_{r,0}/\Omega_{m,0}$. Assim, obtemos~\cite{barbara}
\begin{equation}
\label{t_grande_mat_rad}
a(t)\approx\left(\frac{3}{2}\Omega_{m,0}^{1/2}H_{0}t\right)^{2/3}\;.
\end{equation}

Compare as equações (\ref{t_grande_mat_rad}) e (\ref{a_materia_plano}). Como, para um Universo plano composto por matéria, $t_{0}=2/3H_{0}$~\cite{liddle,weinberg,cheng}, é possível perceber que as duas equações são extremamente similares. Portanto, se considerarmos tempos suficientemente grandes, i.e., $t\gg t_{rm}$, um Universo espacialmente plano composto por matéria e radiação, se comportará de forma análoga aquele cuja única componente é matéria~\cite{bari}.\footnote{Nesse caso, não podemos estender essa conclusão para um Universo plano contemplado com todos os constituintes, dado que para fatores de escalas ainda maiores do que $a_{rm}$ o termo referente à $\Lambda$ será relevante.}

\subsection{Universo Composto por Matéria e Constante Cosmológica}
\label{mat_lambda}
Prosseguindo da mesma forma que fizemos para o cenário contendo matéria e radiação, podemos obter o fator de escala de equivalência entre matéria e Constante Cosmológica, {definido por $a_{m\Lambda}$, que será dado por} ~\cite{barbara,bari}
\begin{equation}
\label{igualdade_mat_lambda}
a_{m\Lambda}=\left(\frac{\Omega_{m,0}}{\Omega_{\Lambda,0}}\right)^{1/3}\;.
\end{equation}
Se fizermos uso dos valores da Eq.~(\ref{par_valores}), resulta { 
\begin{equation}
\label{a_coexistencia_mat_lambda}
a_{m\Lambda}=0,77\;.
\end{equation}}
Tendo em vista que o fator de escala de equivalência $a_{rm}$, dado pela Eq.~(\ref{a_coexistencia_mat_rad}), é muito menor do que o fator de escala de equivalência $a_{m\Lambda}$, dado pela Eq.~(\ref{a_coexistencia_mat_lambda}) e como a contribuição da radiação decresce rapidamente (com $a^{-4}$) conforme o Universo se expande, um Universo plano composto por matéria e Constante Cosmológica é uma aproximação muito boa do que se observa atualmente~\cite{weinberg,barbara}. Por conseguinte, podemos reescrever a Eq.~(\ref{friedmann_omega}) para o presente cenário da seguinte forma
\begin{equation}
\label{ed_mat_lamb>0}
\dot{a}=H_{0}\Omega_{\Lambda,0}^{1/2}\left(\frac{a_{m\Lambda}^{3}}{a}+a^{2}\right)^{1/2}\;,
\end{equation}
\noindent onde, por integração direta, temos que
\begin{equation}
\label{resol_mat_lamb>0}
\int_{0}^{a}\left(\frac{a_{m\Lambda}^{3}}{\tilde{a}}+\tilde{a}^{2}\right)^{-1/2}d\tilde{a}=H_{0}\Omega_{\Lambda,0}^{1/2}t\;.
\end{equation}
Por uma substituição do tipo $(\tilde{a}/a_{m\Lambda})^{3/2}\rightarrow a'$, resulta~\cite{barbara}
\begin{equation}
\label{sol_mat_lambda>0}
H_{0}t=\frac{2}{3\Omega_{\Lambda,0}^{1/2}}\ln\left[{\left(\frac{a}{a_{m\Lambda}}\right)^{3/2}+\sqrt{1+\left(\frac{a}{a_{m\Lambda}}\right)^{3}}}\right]\;.
\end{equation}
{Essa solução nos possibilita determinar o instante de tempo, $t_{m\Lambda}$, no qual as contribuições de matéria e Constante Cosmológica são equivalentes}, bem como viabiliza a análise assintótica que desejamos. 

Primeiramente, fazendo $a=a_{m\Lambda}$ na Eq.~(\ref{sol_mat_lambda>0}), resulta~\cite{barbara}
\begin{equation}
\label{idade_coexist_materia_lambda}
t_{m\Lambda}=\frac{2}{3\Omega_{\Lambda,0}^{1/2}}\ln{\left(1+\sqrt{2}\right)}H_{0}^{-1}\;.
\end{equation}
{Usando as equações (\ref{const_hubble}) e (\ref{par_valores})}, teremos que a idade que um Universo plano deve possuir para que haja igual contribuição de matéria e $\Lambda$ é {
\begin{equation}
\label{idade_coexist_materia_lambda2}
t_{m\Lambda}= 10,3\times10^{9} \textrm{\,anos}\;.
\end{equation}
}Note a diferença entre as ordens de grandeza de $t_{rm}$ e $t_{m\Lambda}$. Enquanto que $t_{rm}\propto 10^{4}\textrm{\,anos}$, temos que $t_{m\Lambda}$ é $10^{5}$ vezes maior. Em adição a isso, lembre-se que para um Universo plano, dominado por matéria, podemos obter a idade do Universo através da Eq.~(\ref{idade_materia}), tal que {$t_{0}\approx 9,7\times 10^{9}\textrm{\,anos}$}. Logo, tomando um Universo composto apenas por matéria e Constante Cosmológica, temos uma boa aproximação na descrição da evolução do Universo, bem como no cálculo de sua idade, $t_{0}$~\cite{bari}. 

{Tendo isso em vista, a idade do Universo pode ser obtida fazendo $t=t_{0}$ na Eq.~(\ref{sol_mat_lambda>0}) e, lembrando que $a(t_{0})=1$, logo
\begin{equation}
\label{idade_materia_lambda}
t_{0}=\frac{2}{3\sqrt{\Omega_{\Lambda,0}}}\ln{\left[\frac{\sqrt{\Omega_{\Lambda,0}}+1}{\sqrt{1-\Omega_{\Lambda,0}}}\right]}H_{0}^{-1}\;.
\end{equation}
Usando os valores dados pelas equações (\ref{const_hubble}) e (\ref{par_valores}), resulta 
\begin{equation}
\label{idade_materia_lambda2}
t_{0} = 13,8\times10^{9}\textrm{\,anos}\;.
\end{equation}
Portanto,  a contribuição da radiação para o cálculo da idade do Universo é muito pequena. Considerando-se a radiação, a idade do Universo sofre uma  variação de apenas algumas partes por milhão~\cite{barbara}. Este resultado é verificado na \autoref{graf_mat_lambda_numerico} onde comparamos os resultados obtidos desconsiderando a contribuição da radiação com àqueles considerando as três contribuições. Vemos que os resultados basicamente coincidem.}

{Além disso, podemos comparar a idade  quando há a equivalência entre matéria e $\Lambda$, $t_{m\Lambda}$, dada pela Eq.~(\ref{idade_coexist_materia_lambda2}), e a idade do Universo $t_0$, obtida quando considerarmos a presença de matéria e Constante Cosmológica, dada pela Eq.~(\ref{idade_materia_lambda2}). Nota-se que, embora ambas são da mesma ordem de grandeza, $t\propto 10^{9}$ anos, há uma diferença, não desprezível, de aproximadamente $3,5\times10^{9}$ anos. Analisemos agora como a solução dada pela Eq.~(\ref{sol_mat_lambda>0}) se comporta em seus limites assintóticos $a\ll a_{m\Lambda}$ e $a\gg a_{m\Lambda}$.}
\subsubsection{Comportamento inicial ($a\ll a_{m\Lambda}$)}
\label{t<<t_mLambda}
Considerando $a\ll a_{m\Lambda}$, podemos utilizar a seguinte expansão
\begin{widetext}
\begin{equation}
\label{exp_serie_ln}
\ln{\left[1+\left(\frac{a}{a_{m\Lambda}}\right)^{3/2}\right]}\approx \left(\frac{a}{a_{m\Lambda}}\right)^{3/2}-\frac{1}{2}\left[\left(\frac{a}{a_{m\Lambda}}\right)^{3/2}\right]^{2}\approx \left(\frac{a}{a_{m\Lambda}}\right)^{3/2}\;.
\end{equation}
\end{widetext}
Substituindo a Eq.~(\ref{exp_serie_ln}) na Eq.~(\ref{sol_mat_lambda>0}) e usando $a_{m\Lambda}^{3}=\Omega_{m,0}/\Omega_{\Lambda,0}$, resulta~\cite{barbara}
\begin{equation}
\label{t_peq_mat_lambda}
a(t)\approx \left(\frac{3}{2}\Omega_{m,0}^{1/2}H_{0}t\right)^{2/3}\;.
\end{equation}
À vista disso, temos que um Universo plano composto por matéria e $\Lambda$ se comportará, para $t\ll t_{m\Lambda}$, como um Universo espacialmente plano dominado por matéria~\cite{bari}. Podemos notar isso através da simples comparação entre as equações (\ref{a_materia_plano}) e (\ref{t_peq_mat_lambda}). 

Repare que esse resultado é idêntico ao obtido na Eq.~(\ref{t_grande_mat_rad}), {onde consideramos $t\gg t_{rm}$ e um Universo plano composto por matéria e radiação}. Logo, percebe-se que, de fato, o período onde a componente referente à matéria dominou sobre as outras está entre o período de dominância dessas.
\subsubsection{Comportamento final ($a\gg a_{m\Lambda}$)}
\label{t>>t_mLambda}
A solução da equação de Friedmann (\ref{sol_mat_lambda>0}) no limite onde $a\gg a_{m\Lambda}$ é dada por
\begin{equation}
H_{0}t\approx \frac{2}{3\Omega_{\Lambda,0}^{1/2}}\ln{\left[2\left(\frac{a}{a_{m\Lambda}}\right)^{3/2}\right]}\approx \frac{1}{\Omega_{\Lambda,0}^{1/2}}\ln{\left[\frac{a}{a_{m\Lambda}}\right]}\;.
\end{equation}
Consequentemente, temos que o comportamento do fator de escala, para um Universo plano composto por matéria e Constante Cosmológica, no limite onde $t\gg t_{m\Lambda}$, é~\cite{barbara} 
\begin{equation}
\label{t_grande_mat_lambda}
a(t)\approx a_{m\Lambda}\exp{\left(\Omega_{\Lambda,0}^{1/2}H_{0}t\right)}\;.
\end{equation}

Perceba a similaridade da Eq.~(\ref{t_grande_mat_lambda}) com a Eq.~(\ref{a_lambda}). Dessa similaridade, podemos notar que um Universo plano composto por matéria e Constante Cosmológica irá se comportar de forma análoga a um Universo plano composto apenas pela Constante Cosmológica, se considerarmos $t\gg t_{m\Lambda}$~\cite{bari}.

\begin{figure}[t]
\centering
\begin{minipage}{8.5cm}
\centering
\includegraphics[width=8.5cm]{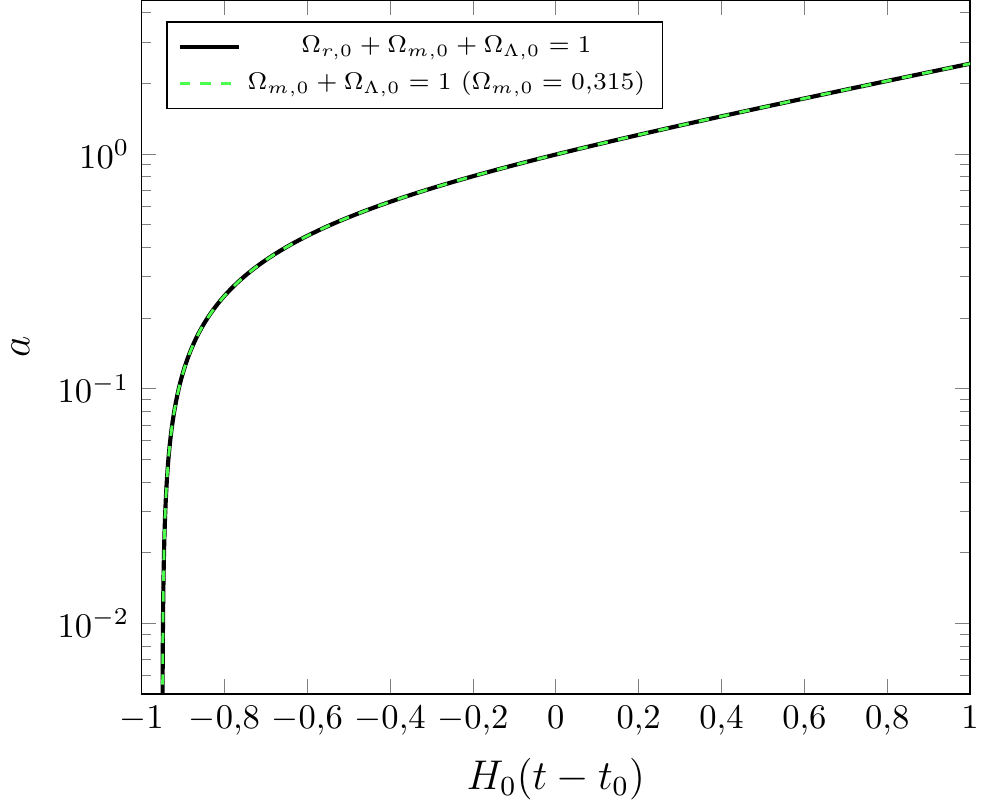}
\caption{{Comparação entre o fator de escala em função do tempo, para um Universo espacialmente plano, composto por matéria e Constante Cosmológica, e aquele obtido com os três constituintes, onde utilizamos $\Omega_{r,0}=2,47\times10^{-5}~h^{2}$, $\Omega_{m,0}=0,315\pm0,007$ e $\Omega_{\Lambda,0}=0,685\pm0,007$.}}
\label{graf_mat_lambda_numerico}
\end{minipage}
\end{figure}

É pertinente descrevermos graficamente o comportamento do fator de escala para qualquer valor temporal e não somente em seus limites assintóticos. Isto posto, utilizamos o método numérico Runge-Kutta de resolução de equações diferenciais a fim de solucionarmos a Eq.~(\ref{ed_mat_lamb>0}), obtendo o comportamento do fator de escala sem restrições. O resultado obtido pode ser visualizado na \autoref{graf_mat_lambda_numerico}. Note que há uma mudança de comportamento nos entornos de $t_{m\Lambda}$. {Tal comportamento deve-se ao fato que um Universo plano composto por matéria e $\Lambda$ se comporta inicialmente como aquele cuja única componente é a matéria e, em um instante posterior ao tempo de equivalência, alterna para um comportamento similar ao de um Universo composto apenas pela Constante Cosmológica\footnote{Estamos utilizando escala logarítimica no eixo dos fatores de escala com o intuito de melhorar a vizualização da mudança de comportamento do fator de escala para $t\ll t_{m\Lambda}$ e $t\gg t_{m\Lambda}$.}. }
\section{Universo composto por Radiação, Matéria e Constante Cosmológica}
\label{completo}
Por fim, expandimos nossas análises ao  cenário mais real, ou seja, um Universo espacialmente plano composto por radiação, matéria e pela Constante Cosmológica. Para esse caso não podemos obter soluções assintóticas como  para os casos apresentados nas Seções \ref{mat_rad} e \ref{mat_lambda}. Todavia, podemos obter a solução numéria utilizando o método Runge-Kutta e tirar algumas conclusões que na verdade, como veremos, são reafirmações do que construímos durante nossas discussões anteriores. { A solução que obtivemos numericamente é apresentada na \autoref{graf_mat_lambda_numerico}, onde fica clara a similaridade com a solução obtida {e} apresentada na Seção anterior. Como já discutido anteriormente, tal resultado deve-se ao fato de que o período no qual a radiação possui contribuição relevante é extremamente diminuto com relação ao período onde matéria e Constante Cosmológica dominam.} 

\begin{figure}[t]
\centering
\begin{minipage}{8.5cm}
\centering
\includegraphics[width=8.5cm]{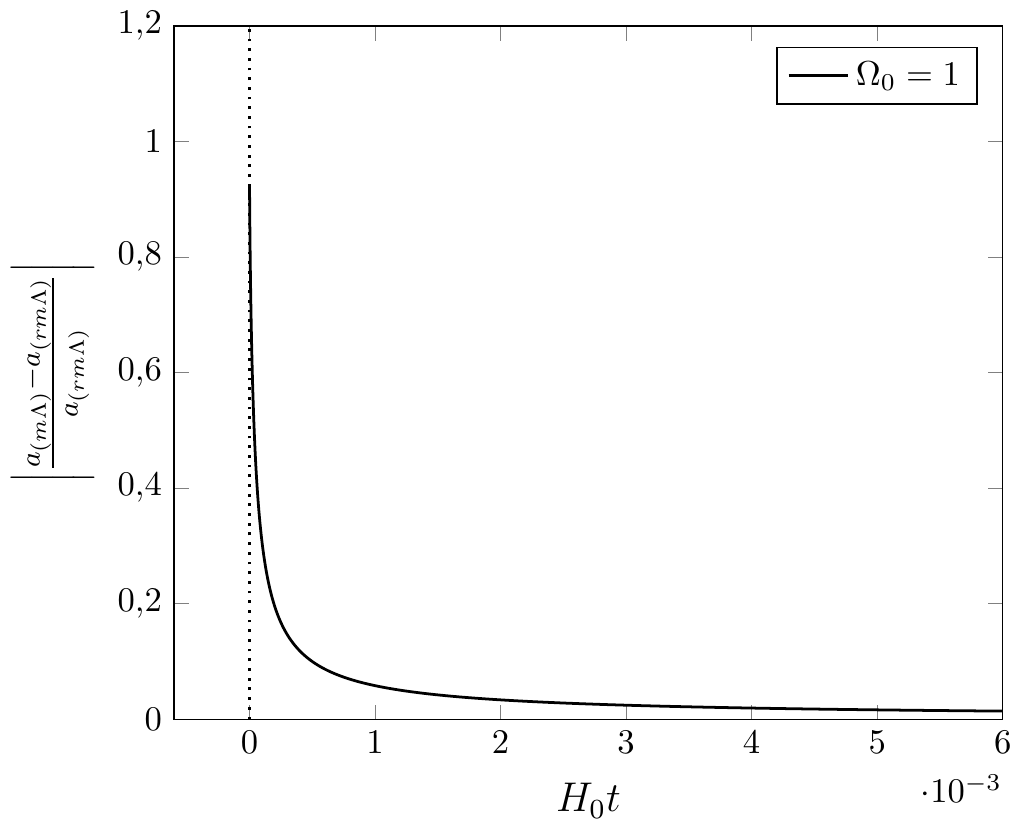}
\caption{Dependência temporal da grandeza definida por $|(a_{(m\Lambda)} - a_{(rm\Lambda)})/a_{(rm\Lambda)}|$ para um Universo plano ($k=0$).}
\label{comparativo}
\end{minipage}
\end{figure}

{ Na \autoref{comparativo} apresentamos a dependência temporal da grandeza definida por $|(a_{(m\Lambda)} - a_{(rm\Lambda)})/a_{(rm\Lambda)}|$, a qual nos permite estimar o impacto da radiação na evolução do Universo. Podemos perceber que, para pequenos valores de tempo, a grandeza cresce fortemente, enquanto para grandes intervalos de tempo ela tende a zero. É facil entender estes resultados,  levando-se em consideração que a contribuição da radiação é extremamente significativa para valores de tempo menores do que $t_{rm}$. Todavia, rapidamente a diferença $a_{(m\Lambda)} - a_{(rm\Lambda)}$ tende à zero,  pois para $t\gg t_{rm}$ a contribuição da radiação se torna insignificante rapidamente, fazendo com que um Universo contendo as três componentes se comporte como um Universo contendo apenas matéria e $\Lambda$.}
\section{Conclusão}
\label{conclusao}
Neste artigo abordamos tópicos essenciais  à Cosmologia, necessários  para obtermos a equação mais importante de nossa análise: a equação de Friedmann. Através da referida equação e levando-se em consideração o cenário especulativo onde a única composição é devida a matéria, analisamos na Seção~\ref{curvatura} como cada tipo de geometria (hiperbólica, esférica e plana) influenciam no destino final do Cosmos. 

Feito isso, utilizamos dos modelos especulativos mais elementares associados a uma geometria plana, expostos nas Seções~\ref{materia},~\ref{radiaçao} e~\ref{lambda}, para demonstrar que, de fato, a aceleração da expansão cósmica é devida à presença da Constante Cosmológica não nula. Verificamos isso observando que a taxa de expansão, tanto para um Universo plano contendo apenas matéria quanto para um Universo plano contendo apenas radiação, possui dependência do tipo $H(t)\propto t^{-1}$, i.e., uma taxa de expansão que decresce com o tempo, não refletindo a aceleração observada~\cite{perlmutter,perlmutter2,riess}. Por sua vez, a taxa de expansão para o cenário composto por $\Lambda$ e com $k=0$ é dada por $H(t)=H_{0}$, i.e., uma taxa de expansão que permanece constante durante toda a evolução do Cosmos e, portanto, podendo explicar a aceleração observada~\cite{perlmutter,perlmutter2,riess}.

Refinando nossas análises, avançamos para cenários compostos por duas componentes, onde nos restringimos aos casos onde a composição é dada, primeiramente, por matéria e radiação e, posteriormente por matéria e Constante Cosmológica. Através da consideração de um Universo plano composto por matéria e radiação, foi possível calcular a idade necessária que o Universo deve ter para que essas duas componentes contribuam igualmente para o conteúdo material do Universo. Tal idade é de aproximadamente {$22.951\textrm{\,anos}$} e marca, para esse cenário, a transição entre um período no qual a radiação domina sobre a matéria para um período onde a matéria domina sobre a radiação. Para o cenário composto por matéria e Constante Cosmológica,  extraímos que o Universo deve possuir aproximadamente {$10,3\times10^{9}\textrm{\,anos}$} para que essas componentes contribuam de forma igualitária para com a composição deste cenário, com este instante de tempo marcando a transição de um período de dominância de matéria para um período de dominância de Constante Cosmológica.

Podemos subdividir a história do Cosmos em três etapas principais. Em uma primeira etapa, que dura cerca de {$23.000\textrm{\,anos}$}, temos que a radiação possui dominância sobre as outras componentes. Logo após, temos a etapa onde a matéria domina sobre as outras componentes. Por fim, quando o Universo possui cerca de {$10,3\times10^{9}\textrm{\,anos}$} há a {transição} para a última etapa onde a Constante Cosmológica domina sobre as outras componentes. Podemos observar, qualitativamente, esta sequência de períodos através da \autoref{evolucao_rho}. 

Dado a grande diferença na ordem de grandeza das idades de equivalência dos dois cenários mencionados anteriormente, constatamos que {considerar como constituintes básicos a matéria e a Constante Cosmológica é uma boa aproximação para se descrever a maior parte da evolução do Universo}. Essa afirmação se torna ainda mais evidente quando a grandeza $|(a_{(m\Lambda)} - a_{(rm\Lambda)})/a_{(rm\Lambda)}|$ a qual tende rapidamente a zero, demonstrando que o fator de escala do cenário completo se aproxima rapidamente do fator de escala do cenário contendo matéria e $\Lambda$. Além disso, considerando a grande proximidade desse cenário para com o cenário completo, foi possível calcular a idade  do Universo como sendo de {$13,8\times10^{9}\textrm{\,anos}$, o qual concorda perfeitamente com o valor obtido a partir dos  dados mais recentes~\cite{pdg2020}}.

{Por fim, esperamos que este artigo contribua para que cada vez mais estudantes de graduação e professores de Física do ensino médio compreendam os conceitos básicos presentes em Cosmologia e, em particular, entendam a evolução do Universo a partir da equação de Friedmann e da modelagem de sua constituição. Recomendamos fortemente ao leitor interessado que  busque complementar as informações apresentadas neste artigo através do estudo das referências \cite{ioav,ioav2,soares,pordeus,froes,viglioni}, publicadas em edições anteriores da Revista Brasileira de Ensino de Física, assim como do livro - texto da Ref.~\cite{ronaldo}.} 

\section*{Agradecimentos}
 {Agradecemos aos revisores pelos importantes comentários que possibilitaram a qualificação do manuscrito.} Este trabalho foi parcialmente financiado pela CNPq,FAPERGS e INCT-FNA (processo número 464898/2014-5).
\section*{Material Suplementar}
O seguinte material suplementar está disponível online.
\begin{itemize}
\item Apêndice A: Da Relatividade Geral à Equação de Friedmann.\\
\item Apêndice B: A Equação de Fluido.\\
\item Apêndice C: A Idade do Universo para $k\neq 0$ e $\Omega_{0}=\Omega_{m,0}$.
\end{itemize}


\newpage
\qquad
\newpage
\begin{widetext}
\section*{Apêndice}
\subsection{Da Relatividade Geral à Equação de Friedmann}
\label{apendice_friedmann}
A métrica que descreve um Universo homogêneo e isotrópico, como vimos na \autoref{RG_EF}, é a métrica de Robertson-Walker~\cite{robertson1,robertson2,robertson3,walker}, cujos coeficientes métricos são
\begin{eqnarray*}
g_{00}=1\;,\;\;\;\;\;\;g_{11}=-a(t)^{2}\frac{1}{1-kr^{2}/R_{0}^{2}}\;,\;\;\;\;\;\;g_{22}=-a(t)^{2}r^{2}\;,\;\;\;\;\;\;g_{33}=-a(t)^{2}r^{2}\sin^{2}\theta\;.
\end{eqnarray*}
Esta métrica  pode ser escrita de uma forma mais compacta
\begin{eqnarray}
\label{metrica_compacta}
g_{00}=1\;\;\;\;\;,\;\;\;\;\;g_{\mu\lambda}=\left\{\begin{array}{rcl}
-a(t)^{2}\tilde{g}_{\mu\lambda}&\mbox{se}&\mu=\lambda\\
0&\mbox{se}&\mu\neq\lambda
\end{array}\right..
\end{eqnarray}

A partir da Eq. (\ref{metrica_compacta}), podemos obter os símbolos de Christoffel, dados por:
\begin{equation}
\label{simbolos}
\Gamma^{\rho}_{\mu\lambda}=\frac{1}{2}g^{\rho\sigma}\left(\partial_{\lambda}g_{\mu\sigma}+\partial_{\mu}g_{\lambda\sigma}-\partial_{\sigma}g_{\mu\lambda}\right)\;.
\end{equation}
Como consequência, temos que:
\begin{eqnarray*}
\Gamma^{0}_{00}=\frac{1}{2}g^{0\sigma}\left(\partial_{0}g_{0\sigma}+\partial_{0}g_{0\sigma}-\partial_{\sigma}g_{00}\right)\;.
\end{eqnarray*}
Ao inserirmos os coeficientes métricos, dados na Eq. (\ref{metrica_compacta}), obtemos
\begin{eqnarray}
\label{simbolos_1}
\Gamma^{0}_{00}=0\;.
\end{eqnarray}
A Eq.~(\ref{simbolos}) também nos diz que
\begin{eqnarray*}
\Gamma^{0}_{\mu0}=\frac{1}{2}g^{0\sigma}\left(\partial_{0}g_{\mu\sigma}+\partial_{\mu}g_{0\sigma}-\partial_{\sigma}g_{\mu0}\right)\;,
\end{eqnarray*}
onde, ao levarmos em consideração a simetria $\Gamma^{\rho}_{\mu\lambda}=\Gamma^{\rho}_{\lambda\mu}$ e a Eq.~(\ref{metrica_compacta}), resulta
\begin{eqnarray}
\label{simbolos_2}
\Gamma^{0}_{\mu0}=\Gamma^{0}_{0\mu}=0\;.
\end{eqnarray}
Os símbolos de Christoffel não nulos são obtidos da mesma maneira. Da Eq. (\ref{simbolos}) temos que
\begin{eqnarray*}
\Gamma^{0}_{\mu\lambda}&=&\frac{1}{2}g^{0\sigma}\left(\partial_{\lambda}g_{\mu\sigma}+\partial_{\mu}g_{\lambda\sigma}-\partial_{\sigma}g_{\mu\lambda}\right)\\
&=&\frac{1}{2}g^{00}\left(\partial_{\lambda}g_{\mu0}+\partial_{\mu}g_{\lambda0}-\partial_{0}g_{\mu\lambda}\right)\\
&=&-\frac{1}{2}\partial_{0}g_{\mu\lambda}\;,
\end{eqnarray*}
onde, utilizando a Eq.~(\ref{metrica_compacta}), e levando em consideração que $\tilde{g}_{\mu\lambda}$ não possui dependência temporal, obtemos
\begin{equation}
\label{simbolos_3}
\Gamma^{0}_{\mu\lambda}=\Gamma^{0}_{\lambda\mu}=\frac{1}{c}a(t)\dot{a}(t)\tilde{g}_{\mu\lambda}\;.
\end{equation}
Além disso, temos também que
\begin{eqnarray*}
\Gamma^{\rho}_{0\lambda}&=&\frac{1}{2}g^{\rho\sigma}\left(\partial_{\lambda}g_{0\sigma}+\partial_{0}g_{\lambda\sigma}-\partial_{\sigma}g_{0\lambda}\right)\\
&=&\frac{1}{2}g^{\rho\rho}\left(\partial_{\lambda}g_{0\rho}+\partial_{0}g_{\lambda\rho}-\partial_{\rho}g_{0\lambda}\right)\\
&=&\frac{1}{2}g^{\rho\rho}\partial_{0}g_{\lambda\rho}\;,
\end{eqnarray*}
de forma que, ao utilizarmos a Eq.~(\ref{metrica_compacta}), teremos
\begin{eqnarray*}
\Gamma^{\rho}_{0\lambda}&=&\frac{1}{2}\left(-\frac{\tilde{g}^{\rho\rho}}{a(t)^{2}}\right)\partial_{0}\left(-a(t)^{2}\tilde{g}_{\lambda\rho}\right)\\
&=&\frac{1}{2c}\frac{\tilde{g}^{\rho\rho}}{a(t)^{2}}2a(t)\dot{a}(t)\tilde{g}_{\lambda\rho}\;.
\end{eqnarray*}
Consequentemente,
\begin{equation}
\label{simbolos_4}
\Gamma^{\rho}_{0\lambda}=\Gamma^{\rho}_{\lambda0}=\frac{1}{c}\frac{\dot{a}(t)}{a(t)}\delta^{\rho}_{\lambda}\;.
\end{equation}
Por fim, definimos
\begin{equation}
\label{simbolos_5}
\Gamma^{\rho}_{\mu\lambda}=\frac{1}{2}\tilde{g}^{\rho\sigma}\left(\partial_{\lambda}\tilde{g}_{\mu\sigma}+\partial_{\mu}\tilde{g}_{\lambda\sigma}-\partial_{\sigma}\tilde{g}_{\mu\lambda}\right)\equiv\tilde{\Gamma}^{\rho}_{\mu\lambda}\;.
\end{equation}

{Partindo da premissa de que o Universo é homogêneo e isotrópico, devemos ter que a sua curvatura deve ser constante. Consequentemente, os tensores que descrevem a curvatura não devem possuir derivadas do tensor métrico, $g_{\mu\nu}$, mas devem depender do tensor métrico em si. Uma escolha do tensor de curvatura que satisfaz isso é~\cite{cheng} 
\begin{equation}
\label{curv_1}
R^{\sigma}_{\mu\nu\lambda}g_{\sigma\rho}=R_{\mu\nu\lambda\rho}=\frac{k}{R^{2}_{0}}\left(g_{\mu\lambda}g_{\nu\rho}-g_{\mu\rho}g_{\nu\lambda}\right)\;,
\end{equation}
onde $k$ é o termo de curvatura e $R_{0}$ é o raio de curvatura. Se desconsiderarmos a variação temporal, podemos reescrever a Eq.~(\ref{curv_1}) como
\begin{equation}
\label{curv_2}
R_{ijkl}=\frac{k}{R^{2}_{0}}\left(g_{ik}g_{jl}-g_{il}g_{jk}\right)\;.
\end{equation}
Da Eq. (\ref{curv_2}), podemos extrair a parte espacial do tensor de Ricci da seguinte forma
\begin{eqnarray*}
\label{ricci_1}
R_{jl}=R_{ijkl}g^{ik}=\frac{k}{R^{2}_{0}}\left(g_{ik}g_{jl}-g_{il}g_{jk}\right)g^{ik}\;,
\end{eqnarray*}
mas $g_{jk}g^{ik}=\delta^{i}_{j}$, logo
\begin{equation}
\label{ricci_2}
R_{jl}=\frac{k}{R^{2}_{0}}\left(3g_{jl}-g_{il}\delta^{i}_{j}\right)=2\frac{k}{R^{2}_{0}}g_{jl}\;
\end{equation}
é a parte espacial do tensor de Ricci.}

{De forma geral, o tensor de Ricci é dado por~\cite{das,kip}}
\begin{equation}
\label{eqricci_1}
R_{\mu\nu}=\partial_{\lambda}\Gamma^{\lambda}_{\mu\nu}-\partial_{\nu}\Gamma^{\lambda}_{\mu\lambda}+\Gamma^{\rho}_{\mu\nu}\Gamma^{\lambda}_{\rho\lambda}-\Gamma^{\rho}_{\mu\lambda}\Gamma^{\lambda}_{\rho\nu}\;,
\end{equation}
{e já que temos os símbolos de Christoffel necessários, podemos obter os termos não nulos do tensor de Ricci.} Dessa forma, temos de (\ref{eqricci_1}) que
\begin{eqnarray*}
R_{00}=\partial_{\lambda}\Gamma^{\lambda}_{00}-\partial_{0}\Gamma^{\lambda}_{0\lambda}+\Gamma^{\rho}_{00}\Gamma^{\lambda}_{\rho\lambda}-\Gamma^{\rho}_{0\lambda}\Gamma^{\lambda}_{\rho0}\;,
\end{eqnarray*}
que, utilizando os símbolos de Christoffel calculados anteriormente, resulta
\begin{eqnarray*}
R_{00}&=&-\partial_{0}\Gamma^{\lambda}_{0\lambda}-\Gamma^{\rho}_{0\lambda}\Gamma^{\lambda}_{\rho0}\\
&=&-\left(\partial_{0}\Gamma^{0}_{00} + \partial_{0}\Gamma^{1}_{01} + \partial_{0}\Gamma^{2}_{02} + \partial_{0}\Gamma^{3}_{03}\right) - \left(\Gamma^{\rho}_{00}\Gamma^{0}_{\rho0} + \Gamma^{\rho}_{01}\Gamma^{1}_{\rho0} + \Gamma^{\rho}_{02}\Gamma^{2}_{\rho0} + \Gamma^{\rho}_{03}\Gamma^{3}_{\rho0}\right)\\
&=&-3\partial_{0}\left(\frac{1}{c}\frac{\dot{a}(t)}{a(t)}\right)-3\left(\frac{1}{c}\frac{\dot{a}(t)}{a(t)}\right)^{2}\;,
\end{eqnarray*}
logo,
\begin{equation}
\label{ricci_00}
R_{00}=-\frac{3}{c^{2}}\frac{\ddot{a}(t)}{a(t)}\;.
\end{equation}
Podemos ainda obter
\begin{equation}
\label{eqricci_3}
R_{ij}=\tilde{R}_{ij}+\left[\frac{a(t)\ddot{a}(t)}{c^{2}}+2\frac{\dot{a}(t)^{2}}{c^{2}}\right]\tilde{g}_{ij}\;,
\end{equation}
onde $\tilde{R}_{ij}$ é a parte espacial do tensor de Ricci, que é obtida puramente através de $\tilde{g}_{ij}$. Mas, podemos utilizar a Eq.~(\ref{ricci_2}), de forma que, se inserirmos na Eq.~(\ref{eqricci_3}), resulta
\begin{equation}
\label{eqricci_4}
R_{ij}=2\frac{k}{R^{2}_{0}}\tilde{g}_{ij}+\left[\frac{a(t)\ddot{a}(t)}{c^{2}}+2\frac{\dot{a}(t)^{2}}{c^{2}}\right]\tilde{g}_{ij}=\left[2\frac{k}{R^{2}_{0}}+\frac{a(t)\ddot{a}(t)}{c^{2}}+2\frac{\dot{a}(t)^{2}}{c^{2}}\right]\tilde{g}_{ij}\;.
\end{equation}

Na \autoref{RG_EF}, enunciamos as equações de campo de Einstein,
\begin{equation}
\label{eq_campo_1}
R_{\mu\nu}=\frac{8\pi G}{c^{4}}\left(T_{\mu\nu}-\frac{1}{2}g_{\mu\nu}T\right)-\frac{\Lambda}{c^{2}}g_{\mu\nu}\;.
\end{equation}
Podemos, a partir dela, definir um certo tensor de rank 2, $C_{\mu\nu}$, tal que
\begin{equation}
\label{tensor_C}
C_{\mu\nu}=T_{\mu\nu}-\frac{1}{2}g_{\mu\nu}T\;.
\end{equation}
Se fizermos uso do tensor energia-momento, dado pela Eq.~(\ref{tensorfluidoperfeito}),
\begin{equation}
\label{tensorfluidoperfeito_1}
\left[T_{\mu\nu}\right]=\left(\begin{array}{cccc}

                              \rho c^{2} & 0 & 0 & 0\\

                              0          & -P & 0 & 0\\

                              0          & 0 & -P & 0\\

                              0          & 0 & 0 & -P

\end{array}\right)\;,
\end{equation}
de onde extraímos o traço
\begin{equation}
\label{traço}
T=\rho c^{2}-3P\;,
\end{equation}
podemos obter os termos do tensor $C_{\mu\nu}$, através da Eq. (\ref{tensor_C}). Temos então que
\begin{eqnarray*}
C_{00}&=&T_{00}-\frac{1}{2}g_{00}\left(\rho c^{2}-3P\right)\\
&=&\rho c^{2}-\frac{1}{2}\left(\rho c^{2}-3P\right)\;.
\end{eqnarray*}
Consequentemente, conclui-se que
\begin{equation}
\label{C_00}
C_{00}=\frac{1}{2}\left(\rho c^{2}+3P\right)\;.
\end{equation}
Além disso, temos também
\begin{eqnarray*}
C_{ij}&=&T_{ij}-\frac{1}{2}g_{ij}\left(\rho c^{2}-3P\right)\\
&=&-Pg_{ij}-\frac{1}{2}g_{ij}\left(\rho c^{2}-3P\right)\\
&=&-Pg_{ij}-\frac{1}{2}g_{ij}\left(\rho c^{2}-3P\right)\\
&=&-\frac{1}{2}\left(\rho c^{2}-P\right)g_{ij}\;.
\end{eqnarray*}
Inserindo a Eq.~(\ref{metrica_compacta}), obtemos
\begin{equation}
\label{C_ij}
C_{ij}=\frac{1}{2}\left(\rho c^{2}-P\right)a(t)^{2}\tilde{g}_{ij}\;.
\end{equation}

Enfim podemos utilizar os tensores $R_{\mu\nu}$ e $C_{\mu\nu}$ nas equações de campo (\ref{eq_campo_1}). Primeiramente, temos
\begin{eqnarray*}
R_{00}=\frac{8\pi G}{c^{4}}\left(T_{00}-\frac{1}{2}g_{00}T\right)-\frac{\Lambda}{c^{2}}g_{00}=\frac{8\pi G}{c^{4}}C_{00}-\frac{\Lambda}{c^{2}}g_{00}\;,
\end{eqnarray*}
onde, ao utilizarmos as equações (\ref{metrica_compacta}), (\ref{ricci_00}) e (\ref{C_00}), obteremos
\begin{eqnarray*}
-\frac{3}{c^{2}}\frac{\ddot{a}(t)}{a(t)}&=&\frac{8\pi G}{c^{4}}\left[\frac{1}{2}\left(\rho c^{2}+3P\right)\right]-\frac{\Lambda}{c^{2}}\\
&=&\frac{4\pi G}{c^{4}}\left(\rho c^{2}+3P\right)-\frac{\Lambda}{c^{2}}\;,
\end{eqnarray*}
resultando
\begin{equation}
\label{eq_campo_2}
\frac{\ddot{a}(t)}{a(t)}=-\frac{4\pi G}{3c^{2}}\left(\rho c^{2}+3P\right)+\frac{\Lambda}{3}\;.
\end{equation}
Note que, da Relatividade Geral, considerando a homogeneidade e isotropia, obtemos naturalmente a equação de aceleração, introduzida na \autoref{RG_EF}.

Por fim, podemos escrever a Eq.~(\ref{eq_campo_1}) para as demais componentes, i.e.
\begin{eqnarray*}
R_{ij}=\frac{8\pi G}{c^{4}}\left(T_{ij}-\frac{1}{2}g_{ij}T\right)-\frac{\Lambda}{c^{2}}g_{ij}=\frac{8\pi G}{c^{4}}C_{ij}-\frac{\Lambda}{c^{2}}g_{ij}\;,
\end{eqnarray*}
assim, se utilizarmos as equações (\ref{metrica_compacta}), (\ref{eqricci_4}) e (\ref{C_ij}), resulta
\begin{eqnarray*}
\left[2\frac{k}{R^{2}_{0}}+\frac{a(t)\ddot{a}(t)}{c^{2}}+2\frac{\dot{a}(t)^{2}}{c^{2}}\right]\tilde{g}_{ij}&=&\frac{8\pi G}{c^{4}}\left[\frac{1}{2}\left(\rho c^{2}-P\right)a(t)^{2}\tilde{g}_{ij}\right]+\frac{\Lambda a(t)^{2}}{c^{2}}\tilde{g}_{ij}\;,\\
\left[2\frac{k}{R^{2}_{0}}+\frac{a(t)\ddot{a}(t)}{c^{2}}+2\frac{\dot{a}(t)^{2}}{c^{2}}\right]\tilde{g}_{ij}&=&\left[\frac{4\pi G}{c^{4}}\left(\rho c^{2}-P\right)a(t)^{2}+\frac{\Lambda a(t)^{2}}{c^{2}}\right]\tilde{g}_{ij}\;,
\end{eqnarray*}
onde, claramente temos que 
\begin{eqnarray*}
2\frac{k}{R^{2}_{0}}+\frac{a(t)\ddot{a}(t)}{c^{2}}+2\frac{\dot{a}(t)^{2}}{c^{2}}&=&\frac{4\pi G}{c^{4}}\left(\rho c^{2}-P\right)a(t)^{2}+\frac{\Lambda a(t)^{2}}{c^{2}}\;.
\end{eqnarray*}
Rearranjando, obtemos
\begin{eqnarray}
\label{quase_friedmann}
\frac{2}{c^{2}}\left[\frac{\dot{a}(t)}{a(t)}\right]^{2}=\frac{4\pi G}{c^{4}}\left(\rho c^{2}-P\right)-\frac{1}{c^{2}}\frac{\ddot{a}(t)}{a(t)}+\frac{\Lambda}{c^{2}}-\frac{2k}{a(t)^{2}R^{2}_{0}}\;.
\end{eqnarray}
Por fim, podemos inserir a Eq.~(\ref{eq_campo_2}) na Eq.~(\ref{quase_friedmann}), tal que
\begin{eqnarray*}
\frac{2}{c^{2}}\left[\frac{\dot{a}(t)}{a(t)}\right]^{2}=\frac{4\pi G}{c^{4}}\left(\rho c^{2}-P\right)+\frac{4\pi G}{3c^{4}}\left(\rho c^{2}+3P\right)-\frac{\Lambda}{3c^{2}}+\frac{\Lambda}{c^{2}}-\frac{2k}{a(t)^{2}R^{2}_{0}}\;,
\end{eqnarray*}
onde, ao juntarmos os termos em comum e efetuar as devidas simplificações, obtemos
\begin{equation}
\left[\frac{\dot{a}(t)}{a(t)}\right]^{2}=\frac{8\pi G}{3}\rho-\frac{kc^{2}}{a(t)^{2}R^{2}_{0}}+\frac{\Lambda}{3}\;,
\end{equation}
a qual é a  \textbf{equação de Friedmann}.
\newpage

\subsection{A Equação de Fluido}
\label{apendice_fluido}
Podemos contrair o tensor energia-momento dado pela Eq.~(\ref{tensorfluidoperfeito}), tal que $T^{\mu}_{\nu}=g^{\sigma\mu}T_{\sigma\nu}$. Com isso, podemos escrever~\cite{fayyazuddin,cheng}
\begin{equation}
\label{tensor1}
T^{\mu}_{\nu}=\left(\rho c^{2} + P\right)\frac{U^{\mu}U_{\nu}}{c^{2}} - P\delta^{\mu}_{\nu}\;,
\end{equation}
onde, para um referencial onde o fluido cósmico está em repouso, $U_{0}=U^{0}=c$ e $U_{i}=U^{i}=0$~\cite{cheng}. Teremos, portanto
\begin{equation}
\label{00}
T^{0}_{0}=\left(\rho c^{2} + P\right)\frac{U^{0}U_{0}}{c^{2}} - P\delta^{0}_{0}=\rho c^{2}\;,
\end{equation}
bem como
\begin{equation}
\label{ij}
T^{i}_{j}=T^{j}_{i}=\left(\rho c^{2} + P\right)\frac{U^{i}U_{j}}{c^{2}} - P\delta^{i}_{j}=-P\delta^{i}_{j}\;,
\end{equation}
onde usamos da simetria do tensor energia momento~\cite{cheng}. Por fim,
\begin{equation}
\label{i0}
T^{i}_{0}=T^{0}_{i}=\left(\rho c^{2} + P\right)\frac{U^{i}U_{0}}{c^{2}} - P\delta^{i}_{0}=0\;.
\end{equation}

Resulta da conservação de energia que~\cite{fayyazuddin}
\begin{equation}
\label{conservaçao}
\nabla_{\mu}T^{\mu}_{0}=\partial_{\mu}T^{\mu}_{0}+\Gamma^{\mu}_{\mu\lambda}T^{\lambda}_{0}-\Gamma^{\lambda}_{\mu 0}T^{\mu}_{\lambda}=0\;.
\end{equation}
Dessa forma
\begin{equation}
\label{conservaçao2}
\nabla_{\mu}T^{\mu}_{0}=\left(\partial_{0}T^{0}_{0} + \partial_{i}T^{i}_{0}\right) + \left(\Gamma^{0}_{00}T^{0}_{0} +\Gamma^{0}_{0j}T^{j}_{0} + \Gamma^{i}_{i0}T^{0}_{0} + \Gamma^{i}_{ij}T^{j}_{0}\right) -\left (\Gamma^{0}_{00}T^{0}_{0} +\Gamma^{j}_{00}T^{0}_{j} + \Gamma^{0}_{i0}T^{i}_{0} +\Gamma^{j}_{i0}T^{i}_{j}\right)=0\;.
\end{equation}
Utilizando os simbolos de Christoffel obtidos no Apêndice A, bem como as equações~(\ref{00})-(\ref{i0}), resulta
\begin{eqnarray}
\label{conservaçao3}
\nabla_{\mu}T^{\mu}_{0}&=&\partial_{0}T^{0}_{0} + \frac{1}{c}\frac{\dot{a}}{a}\delta^{i}_{i}T^{0}_{0}  - \frac{1}{c}\frac{\dot{a}}{a}\delta^{j}_{i}T^{i}_{j}=0\nonumber\\
&=&\frac{\partial(\rho c^{2})}{\partial(ct)}+ \frac{3}{c}\frac{\dot{a}}{a}\rho c^{2}  + \frac{1}{c}\frac{\dot{a}}{a}P\delta^{j}_{i}\delta^{i}_{j}=0\nonumber\\
&=&c\dot{\rho}+ \frac{3}{c}\frac{\dot{a}}{a}\rho c^{2}  + \frac{3}{c}\frac{\dot{a}}{a}P=0\;.
\end{eqnarray}
Rearranjando a Eq.~(\ref{conservaçao3}) obtemos~\cite{fayyazuddin}
\begin{equation}
\dot{\rho}+\frac{3}{c^{2}}\frac{\dot{a}}{a}\left(\rho c^{2} + P\right)=0\;,
\end{equation}
o qual denomina-se \textbf{equação de fluido}.

\newpage

\subsection{A Idade do Universo para $k\neq 0$ e $\Omega_{0}=\Omega_{m,0}$}
\label{apendice_idade}

\begin{figure}[b!]
\centering
\begin{minipage}{8.5cm}
\centering
\includegraphics[width=8.5cm]{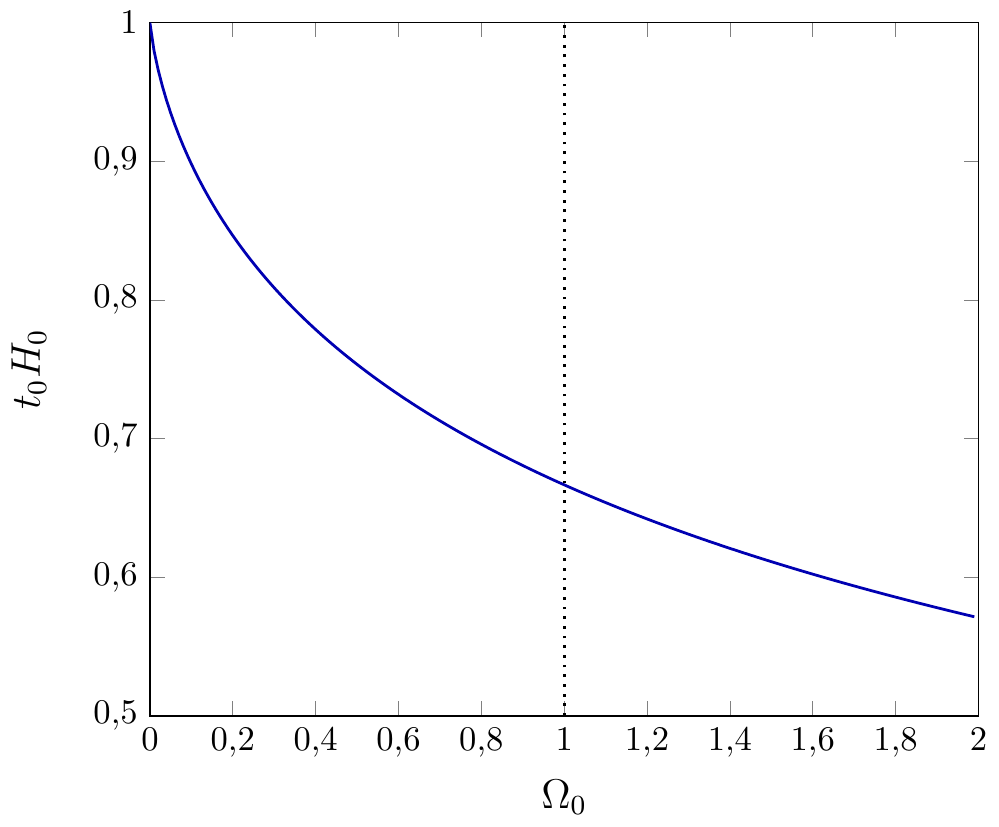}
\caption{Comportamento de $t_{0}H_{0}$ em função do parâmetro de densidade $\Omega_{0}$. A seção da curva do lado esquerdo da curva pontilhada diz respeito à Eq.~(\ref{idade_k=-1}) enquanto que a seção à direita diz respeito à Eq.~(\ref{idade_k=1}).}
\label{evol_idade}
\end{minipage}
\end{figure}


Para obtermos a idade $t_{0}$ de um Universo contendo matéria e $\Omega_{0}>1$ devemos primeiramente encontrar o ângulo de desenvolvimento $\alpha_{0}$ que resulte $a = 1$, visto que estamos levando em consideração que $a(t_{0})=1$. Da Eq.~(\ref{a_materia_curvatura1}) extraímos
\begin{equation}
\label{alpha0}
\alpha_{0}=\arccos{\left(1-2\frac{\Omega_{0}-1}{\Omega_{0}}\right)}\;.
\end{equation}
Com isso, podemos inserir a Eq.~(\ref{alpha0}) na Eq.~(\ref{t_materia_curvatura1}), tal que
\begin{equation}
t_{0}=t(\alpha_{0})=\frac{1}{2H_{0}}\frac{\Omega_{0}}{(\Omega_{0}-1)^{3/2}}\left\{\arccos{\left(1-2\frac{\Omega_{0}-1}{\Omega_{0}}\right)}-\sin{\left[\arccos{\left(1-2\frac{\Omega_{0}-1}{\Omega_{0}}\right)}\right]}\right\}\;,
\end{equation}
mas, como $\sin{(\arccos{x})}=\sqrt{1-x^{2}}$, resulta~\cite{narlikar_intro,weinberg2}
\begin{equation}
\label{idade_k=1}
t_{0}=\frac{1}{2H_{0}}\frac{\Omega_{0}}{(\Omega_{0}-1)^{3/2}}\left[\arccos{\left(\frac{2}{\Omega_{0}}-1\right)}-2\frac{\sqrt{\Omega_{0}-1}}{\Omega_{0}}\right]\;.
\end{equation}

Procedendo da mesma forma para um Universo contendo matéria, mas com $\Omega_{0}<1$, obtemos o ângulo de desenvolvimento $\beta_{0}$ através da Eq.~(\ref{a_materia_curvatura2}) fazendo $a=1$, daí
\begin{equation}
\label{beta0}
\beta_{0}=\arccosh{\left(1+2\frac{1-\Omega_{0}}{\Omega_{0}}\right)}\;.
\end{equation}
Introduzimos então a Eq.~(\ref{beta0}) na Eq.~(\ref{t_materia_curvatura2}) e obtemos
\begin{equation}
t_{0}=t(\beta_{0})=\frac{1}{2H_{0}}\frac{\Omega_{0}}{(1-\Omega_{0})^{3/2}}\left\{\sinh{\left[\arccosh{\left(1+2\frac{1-\Omega_{0}}{\Omega_{0}}\right)}\right]}-\arccosh{\left(1+2\frac{1-\Omega_{0}}{\Omega_{0}}\right)}\right\}\;.
\end{equation}
Utilizando o fato de que $\sinh{(\arccosh{x})}=\sqrt{x^{2}-1}$, temos que a idade de um Universo hiperbólico composto apenas por matéria é~\cite{weinberg2}
\begin{equation}
\label{idade_k=-1}
t_{0}=\frac{1}{2H_{0}}\frac{\Omega_{0}}{(1-\Omega_{0})^{3/2}}\left[2\frac{\sqrt{1-\Omega_{0}}}{\Omega_{0}}-\arccosh{\left(\frac{2}{\Omega_{0}}-1\right)}\right]\;.
\end{equation}


Na \autoref{evol_idade} podemos verificar que, para $\Omega_{0}\rightarrow 1$ obtemos que $t_{0}H_{0}\rightarrow2/3$, i.e., o mesmo que esperamos para um Universo plano composto apenas por matéria (veja a Seção~\ref{materia_plano}). Por sua vez, para $\Omega_{0}\rightarrow 0$ resulta que $t_{0}H_{0}\rightarrow1$, a idade esperada para um Universo vazio~\cite{bari} (Universo de Milne~\cite{lars}). 

\end{widetext}

\end{document}